\documentclass[11pt,a4paper]{article}
\pdfoutput=1
\usepackage{jheppub}

\usepackage{multirow}
\usepackage{amsmath,amssymb}
\usepackage{graphicx}
\usepackage{color}
\usepackage{hyperref}
\usepackage{url}
\usepackage{slashed}
\usepackage{subfigure}
\usepackage[usenames,dvipsnames]{xcolor}

\title{Upper bound on the Abelian gauge coupling from asymptotic safety}
\author[a]{Astrid Eichhorn,}
\author[a]{Fleur Versteegen}
\emailAdd{a.eichhorn@thphys.uni-heidelberg.de, f.versteegen@thphys.uni-heidelberg.de}

\affiliation[a]{Institut f\"ur Theoretische Physik, Universit\"at Heidelberg, Philosophenweg 16, 69120 Heidelberg, Germany}
\usepackage{amssymb}
\usepackage{slashed}
\usepackage{graphicx}
\usepackage{graphics}
\usepackage{epsfig}
\usepackage{color}
\usepackage{soul}
\usepackage{tabularx}
\usepackage{amsmath}
\usepackage{amsfonts}
\usepackage{amssymb}
\usepackage{hyperref}
\usepackage{microtype}

\newcommand{\be}{\begin{equation}}
\newcommand{\ee}{\end{equation}}
\newcommand{\bea}{\begin{eqnarray}}
\newcommand{\eea}{\end{eqnarray}}

\title{Upper bound on the Abelian gauge coupling from asymptotic safety
}

\abstract{
We explore the impact of asymptotically safe quantum gravity on the Abelian gauge coupling in a model including a charged scalar, confirming indications that asymptotically safe quantum fluctuations of gravity could trigger a power-law running towards a free fixed point for the gauge coupling above the Planck scale. Simultaneously, quantum gravity fluctuations balance against matter fluctuations to generate an interacting fixed point, which acts as a boundary of the basin of attraction of the free fixed point. This enforces an upper bound on the infrared value of the Abelian gauge coupling. In the regime of gravity couplings which in our approximation also allows for a prediction of the top quark and Higgs mass close to the experimental value \cite{Eichhorn:2017ylw}, we obtain an upper bound approximately 35 $\%$ above the infrared value of the hypercharge coupling in the Standard Model.
}

\begin{document}
\maketitle

\section{Introduction: Quantum-gravity induced UV completion for scalar QED}
The vacuum in scalar Quantum Electrodynamics, and more generally in Abelian gauge theories has screening properties. This results in a beta function with a positive coefficient of the leading perturbative term, translating into a running coupling that increases as a function of the
energy scale. A perturbative treatment is inapplicable beyond a certain scale, but nonperturbative studies support the conclusion that scalar QED is trivial. Hence, the limit of arbitrarily high momentum scales cannot be reached while keeping the coupling finite in the infrared (IR), \cite{Baig:1994bw}. Thus, a finite IR value of the coupling appears to require the introduction of new physics at high scales.
For the Standard Model, the Landau pole that is the perturbative signature connected to the nonperturbative triviality problem, occurs beyond the Planck scale. This already suggests that a solution to the problem could lie in the inclusion of quantum gravitational degrees of freedom.  
We focus on an asymptotically safe model of quantum gravity \cite{Weinberg:1980gg}.
Based on the groundbreaking work of Reuter \cite{Reuter:1996cp}, compelling evidence for the existence of an asymptotically safe fixed point in pure gravity \cite{ASgrav} as well as with matter \cite{Dona:2013qba,Eichhorn:2016vvy,Meibohm:2015twa,Dona:2015tnf,Biemans:2017zca} has been discovered, for reviews see, e.g., \cite{ASreviews}. This fixed point provides a UV completion for a quantum field theory of the metric. Here, we extend these studies by exploring the possibility of an asymptotically safe, quantum-gravity induced UV completion for Abelian gauge theories, see also \cite{Harst:2011zx,Christiansen:2017gtg}.

We restrict ourselves to the canonical interactions for scalar QED coupled to Einstein-Hilbert gravity. Our study includes the Abelian gauge field, a complex scalar, as well as metric fluctuations, for which we keep a general background gauge fixing term with two gauge parameters $\alpha, \beta$. 
 Specifically,
we explore whether quantum gravity can induce an asymptotically free fixed point for the gauge coupling as well as a second,
asymptotically safe one, which enhances the predictive power of the asymptotic safety scenario: If quantum fluctuations of gravity cause the gauge coupling to become an irrelevant direction at an interacting fixed point, this renders its low-energy value predictable, since the fixed point is IR attractive for that coupling. Crucially, this induces an upper bound on the viable IR values of the gauge coupling: Values above that upper bound are shielded from the basin of attraction of the free fixed point by the IR attractive interacting fixed point. Thus, only values below the lower bound can be reached from an ultraviolet complete microscopic model. For all other cases, new physics must necessarily exist.

 This paper is structured as follows: In Sec.~\ref{sec:synposis} we explain the fixed-point structure underlying the upper bound on the gauge coupling.  Sec.~\ref{sec:flows} contains the implications of our results for the Abelian gauge coupling of the Standard Model, i.e., the hypercharge coupling. These two sections are self-contained and can be read without Sec.~\ref{sec:FRGbetas}, which details the techniques underlying our calculation and provides an overview of our results.
In the final section, Sec.~\ref{sec:outlook}, we consider the possible implications of our result together with that in \cite{Eichhorn:2017ylw} for the perspective of a predictive UV completion of the Standard Model. We highlight the potential for the existence of a microscopic model including quantum gravity degrees of freedom and fixing several of the free parameters of the Standard Model. The appendix contains a comprehensive analysis of the gauge dependence of our result, the reasoning underlying a prefered choice of gauge and further technical details.

\section{Synopsis: Upper bound on the gauge coupling from asymptotic safety}\label{sec:synposis}

\begin{figure}[!t]
\begin{center}
\includegraphics[width=0.6\linewidth]{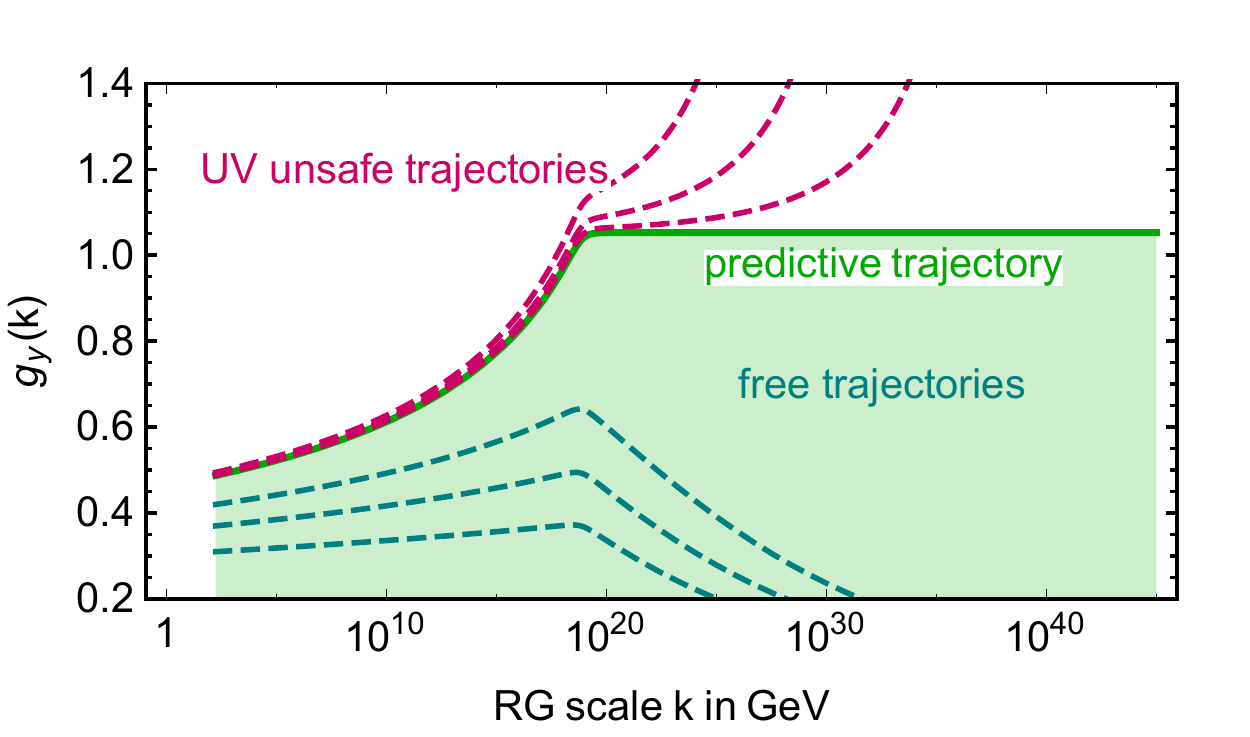}
\end{center}
\caption{\label{fig:upperboundgy} We show the upper bound on the IR value of $g_Y$ under the impact of running gravitational couplings, which show a rapid transition from a classical regime below the Planck scale to a fixed-point regime above the Planck scale.}
\end{figure}

Several couplings are marginally irrelevant in the Standard Model. This includes the Abelian gauge coupling, the Higgs quartic coupling and the Yukawa couplings. Physically, marginal irrelevance means a growth of the coupling towards the ultraviolet. 
Asymptotically safe quantum gravity adds a contribution to the beta function of all matter couplings that is linear in the matter coupling   \cite{Narain:2009fy,Zanusso:2009bs,Daum:2009dn,Vacca:2010mj,Harst:2011zx,Folkerts:2011jz,Eichhorn:2011pc,Eichhorn:2012va,Oda:2015sma,Meibohm:2016mkp,Eichhorn:2016esv,Eichhorn:2016vvy,Christiansen:2017gtg,Hamada:2017rvn,Eichhorn:2017eht,
Eichhorn:2017ylw}. If the sign of the gravitational contribution is negative, a fundamental change is triggered in the high-energy behavior of the corresponding coupling: The UV-repulsive free fixed point is turned into a UV attractive fixed point.  This effect kicks in at the Planck scale, when quantum gravity fluctuations become important. Instead of continuing to grow logarithmically towards the UV, the corresponding matter coupling then decreases and exhibits a power-law running towards asymptotic freedom. 
Thus, finite low-energy values of the coupling become compatible with  ultraviolet completeness. 
 The main question is how large the coupling can become for the quantum-gravity effect to be powerful enough to render the theory well-behaved in the UV. Intuitively, one might imagine that there is a critical strength of the matter coupling, beyond which the quantum fluctuations of matter are just too strong in order for the quantum gravity fluctuations to overwhelm them and trigger a decrease of the coupling strength. Then, at the critical value of the coupling, quantum fluctuations of matter and gravity balance out exactly, such that the coupling remains exactly constant and ceases to run, i.e., it hits an interacting, asymptotically safe fixed point in the UV.
 In fact, this is exactly what happens, and is rooted in the simple structure of the corresponding beta functions:
In the scenario in which quantum gravity can induce asymptotic freedom, quantum fluctuations of matter and gravity enter the beta function of the matter coupling $g$ with opposite signs, schematically
\be
\beta_g= \#_1 g^3 - \#_2 G\,g, 
\ee
with $\#_{1,2}>0$ and $G$ the Newton coupling.
This allows  quantum fluctuations of matter and gravity to balance each other at an interacting fixed point. Crucially, this fixed point must be ultraviolet repulsive and can thus only be connected to a unique infrared value of the coupling. The critical trajectory that emanates from the interacting fixed point divides the range of low-energy values of the coupling into two distinct regimes: 
 For IR values smaller than the critical coupling, metric fluctuations ``win" in the UV and the coupling decreases. Hence, all trajectories which have IR-end-points below the critical value are attracted to the free fixed point towards the UV. These IR values are thus consistent with a UV complete theory.
 On the other hand, for IR values larger than the critical coupling, matter fluctuations remain dominant in the UV and the coupling continues to grow. As the interacting fixed point is UV repulsive, all IR values above the critical value are driven away from it and cannot approach any fixed point in the UV. Thus all IR values above the critical value are screened from a UV-complete regime by the critical trajectory. Fig.~\ref{fig:upperboundgy} showcases the corresponding behavior; details on the corresponding RG flow are explained in Sec.~\ref{sec:flows}.
To summarize, a UV completion that is defined based on the interacting fixed point provides a unique prediction for the IR value of the coupling. Simultaneously, this unique value acts as a strict upper bound for all viable IR values.

\section{Impact of metric fluctuations on the gauge coupling}\label{sec:FRGbetas}
\label{sec:FRGbetas}
\subsection{Functional Renormalization Group setup}

Quantum fluctuations generate all couplings compatible with the symmetries,  even in cases where some of them are set to zero at some scale. Therefore, the RG flow lives in theory space, which is the space of all couplings. A model becomes asymptotically safe if there is an interacting fixed point in theory space, and the IR values of couplings lie at a point which can be reached along an RG trajectory emanating from that fixed point. The fixed point realizes quantum scale invariance, and the departure from scale invariance in the IR is encoded in the values of relevant couplings: These span the UV critical hypersurface of the fixed point, and parameterize all possible IR models that can be reached starting from a UV safe fixed point. Accordingly, predictivity requires the number of relevant couplings to be finite, in order for a finite number of measurements to suffice for pinning down the IR values of the relevant couplings. In turn, these then determine the values of all irrelevant couplings. Those are couplings which span the directions orthogonal to the critical hypersurface at the fixed point. Along those directions, points in the IR \emph{cannot} be connected to the fixed point by an RG trajectory, since the fixed point is UV repulsive in those directions. Therefore, viable RG trajectories lie within the critical hypersurface, which implies that the values of irrelevant couplings at all scales are determined by the values of the relevant couplings. \\

We employ the functional RG (FRG), which is based on a mass-like cutoff term in the Euclidean generating functional, that selectively suppresses quantum fluctuations based on their momenta: For modes with momenta $p^2>k^2$, where $k$ is the RG scale, the cutoff vanishes, and those modes contribute to the effective action at the scale $k$. In contrast, low momentum modes with $p^2<k^2$ are suppressed by the mass-like cutoff term $R_k(p^2)$. Lowering the scale $k$  results in an addition of the contribution of quantum fluctuations with momenta $p^2 \approx k^2$, which results in a scale dependence of the effective dynamics, captured by a scale dependence of the couplings. 
Within the FRG setting, a formally exact equation governs the flow of the effective dynamics \cite{Wetterich:1992yh}
\be
\partial_t \Gamma_k = k\, \partial_k \Gamma_k = \frac{1}{2}{\rm Tr}\left(\Gamma_k^{(2)}+R_k \right)^{-1}\partial_t R_k,\label{eq:Wetterich}
\ee
see also \cite{Morris:1993qb}. For reviews, see \cite{Berges:2000ew, Polonyi:2001se,
Pawlowski:2005xe, Gies:2006wv, Delamotte:2007pf, Rosten:2010vm, Braun:2011pp}. Herein, $\Gamma_k$ is a modified Legendre transform of the $k$-dependent generating functional, which is defined such that it provides the standard effective action when $k\rightarrow 0$. In terms of scale-dependent couplings and field monomials $\mathcal{O}_i$, it can be expanded as
\be
\Gamma_k = \sum_i g_i(k)\mathcal{O}^i,
\ee
i.e., the beta functions of the couplings can be extracted from the Wetterich equation \eqref{eq:Wetterich}
by projecting the right-hand-side onto the corresponding field monomial $\mathcal{O}_i$. $\Gamma_k^{(2)}$ is the second derivative of the effective action with respect to the fields, and accordingly $\left(\Gamma_k^{(2)}+R_k \right)^{-1}$ is the full, IR-regularized propagator of the model. Its eigenvalues are summed/integrated over in the $\rm Tr$, i.e., the flow equation has a one-loop structure. For models with several fields, $\Gamma_k^{(2)}$ becomes a matrix in field space. Accordingly, the contributions to the beta function of a particular coupling are of a one-loop form, with different fields on the internal propagator lines, depending on the coupling in question.

With this tool, interacting fixed points of the Renormalization Group, which can provide a predictive ultraviolet (UV) completion to perturbatively nonrenormalizable models, can be discovered, see, e.g., \cite{Eichhorn:2016hdi,Gies:2013pma,Braun:2010tt,Gies:2009hq,Gies:2003ic}. 

Calculations in the full, infinite dimensional theory space are not possible, and one thus resorts to truncations of  theory space. 
The truncation that we explore consists of an Einstein-Hilbert term, a gauge-fixing term for metric fluctuations, a field-strength term for Abelian gauge bosons, a gauge-fixing term for the Abelian fields, and a gauge- covariant kinetic term for the charged scalar, i.e., 
\bea
\Gamma_k&=& -\frac{1}{16\pi\, G_N}\int d^4x\, \sqrt{g}\left(R- 2\bar{\Lambda} \right) \label{eq:truncation}\\
&{}& +\frac{1}{\alpha\, 32 \pi\, G_N} \int d^4x\, \sqrt{\bar{g}}\bar{g}^{\mu\nu}\left(\bar{D}^{\kappa}h_{\kappa\mu}-\frac{1+\beta}{4}\bar{D}_{\mu}h \right)\cdot \left(\bar{D}^{\lambda}h_{\lambda\nu}-\frac{1+\beta}{4}\bar{D}_{\nu}h \right)\nonumber\\
&{}& + \frac{Z_A}{4}\int d^4x\, \sqrt{g}g^{\mu\nu}g^{\kappa\lambda}F_{\mu\kappa}F_{\nu\lambda}+ \frac{1}{\xi} \int d^4x\, \sqrt{\bar{g}}\left(\bar{g}^{\mu\nu}\bar{D}_{\mu}A_{\nu} \right)^2\nonumber\\
&{}&+Z_{\phi} \int d^4x\, \sqrt{g}g^{\mu\nu}\left(\partial_{\mu} + i\, \bar{\rho}\, A_{\mu} \right)\phi^{\dagger} \left(\partial_{\nu} - i\bar{\rho} A_{\nu} \right)\phi.\nonumber
\eea
where $\alpha, \beta$ denote the gauge parameters for the gravitational part of the action and $\xi$ denotes the gauge parameter for the Abelian field. For the metric fluctuations, we employ a background gauge fixing: After splitting the metric into a background and fluctuations, 
\be
g_{\mu\nu} = \bar{g}_{\mu\nu}+h_{\mu\nu},
\ee
the fluctuations are gauge-fixed with respect to the background. For the evaluation of the flow of the gauge coupling, a flat background
\be
\bar{g}_{\mu\nu} = \delta_{\mu\nu},
\ee
is sufficient and technically preferable.
We have explicitly written out the gauge covariant derivative in the kinetic term for the complex scalar, and by $\bar{D}_{\mu}$ we denote the covariant derivative with respect to the auxiliary background metric $\bar{g}_{\mu\nu}$. As we do not evaluate the running in the gravitational sector in this work, we can neglect the Faddeev-Popov ghosts for both gauge fixing terms. In general, the Faddeev-Popov ghost term for the Abelian gauge fixing cannot be neglected in a setting including gravity, as it contributes to the flow of the background couplings in the gravitational sector, see, e.g., \cite{Daum:2009dn,Dona:2013qba}.

 We work with dimensionless gravitational couplings
\be
G = G_N\, k^{2}, \quad \Lambda = \bar{\Lambda}\,k^{-2},
\ee
and introduce a wave function renormalization $Z_h$ for the metric fluctuations in their kinetic term. The kinetic term, which provides what we losely speaking refer to as the graviton propagator, arises from the expansion of the Einstein-Hilbert term in Eq.~\eqref{eq:truncation} to second order in $h$, followed by a rescaling of the fluctuations by $\sqrt{16\pi\, G_N}$, such that the field $h_{\mu\nu}$ has canonical dimension 1. The anomalous dimension of the fluctuation field $h_{\mu\nu}$ is 
\be
\eta_h = - \partial_t\ln Z_h.
\ee
We choose a theta cutoff with shape function \cite{Litim:2001up}
\be
R_k = Z \left(k^2-p^2 \right)\theta(k^2-p^2),
\ee
multiplied by an appropriate tensor structure for the gauge field and the metric. The tensor structure is chosen such that within the scalar part of the regularized propagator for the fluctuation of each component of $h_{\mu\nu}$, $p^2$ is replaced by $k^2$.
The regulator leads to an additional breaking of gauge invariance, adding another term to the modified Ward-identity, see, e.g., \cite{Reuter:1993kw,Ellwanger:1995qf,DAttanasio:1996tzp,Reuter:1997gx, Litim:1998qi,Freire:2000bq}. In \cite{Reuter:1994sg} the background-field method was employed instead to derive the flow for scalar electrodynamics from the FRG.

Note that we have three choices to extract the running of the gauge coupling, $\rho$. 
It can be extracted from the three-point function, where it provides the momentum-dependent interaction between the charged scalar, its complex conjugate and the gauge field. Alternatively, it can be extracted from the four-point function, where it provides the momentum-independent two-photon-scalar-antiscalar interaction. 
Finally, a rescaling of the gauge field $A_{\mu} \rightarrow \frac{1}{\rho}A_{\mu}$ leads to a form of the action where the running coupling can be read off the gauge field propagator,  as was done, e.g., in \cite{Christiansen:2017gtg}. Gauge invariance of course implies that the running of $\rho$ does not depend on how it is extracted. This is a nontrivial requirement as different diagrams enter the flow equation for the three- and four-point function. Furthermore, the beta function receives contributions from the anomalous dimensions of the field, and since there is one extra gauge field for the four-point coupling compared to the three-point one, the factors of the anomalous dimensions differ. To account for that, we will distinguish the coupling of the running three point vertex from that of the four-point vertex, and write
\bea
\Gamma_k\Big|_{\rho} &=& i \bar{\rho}_3 \int d^4x\, \sqrt{g}g^{\mu\nu}\left(\phi^{\dagger}\partial_{\nu}\phi - (\partial_{\nu}\phi^{\dagger}) \phi\right)A_{\mu} +\bar{\rho}_4^2 \int d^4x\, \sqrt{g}g^{\mu\nu}A_{\mu}A_{\nu}\phi^{\dagger}\phi.
\eea
Normalizing the kinetic terms to canonical prefactors results in the definition of renormalized gauge couplings
\be
\rho_3 = \frac{\bar{\rho}_3}{Z_{\phi}Z_A^{1/2}}, \quad \rho_4^2 = \frac{\bar{\rho}_4^2}{Z_{\phi}Z_A}.
\ee
To obtain a universal beta function at the one-loop level, see the discussion in App.~\ref{sec:beta_app}, it is necessary to include the anomalous dimensions
\be
\eta_{\phi} = - \partial_t \ln Z_{\phi}, \quad \eta_A = - \partial_t \ln Z_A.
\ee
Thus, the two beta functions read
\bea
\label{eq:beta_3}
\beta_{\rho_3}&=& \rho_3 \left(\eta_{\phi} + \frac{1}{2}\eta_A \right)+ ...,\\
\label{eq:beta_4}
\beta_{\rho_4}&=& \rho_4 \left(\frac{\eta_{\phi}}{2} +\frac{\eta_A}{2} \right)+ ...\, ,
\eea
where the dots denote the explicit contributions of the diagrams in Figs.~\ref{fig:betarho3mat}, \ref{fig:betarho4mat}. Upon setting $\rho_4 = \rho_3$, the explicit contributions are $\sim \rho^3$ in the case of $\beta_{\rho_3}$ and $\sim \rho^3$ in the case of $\beta_{\rho_4}$. As $\eta_{\phi/A} \sim \rho_3^2$, we obtain the universal one-loop result $\beta_{\rho} = \frac{1}{48\pi^2} \rho^3$ from all three
ways of extracting the beta function for the case of vanishing gravity and in the perturbative approximation, where $\eta$ terms arising from the scale derivative of the regulator are set to zero.

\subsection{Quantum-gravity contributions in the TT approximation}\label{sec:QGTT}
\begin{figure}[!t]
\begin{minipage}{0.5\linewidth}
\includegraphics[width=0.2\linewidth]{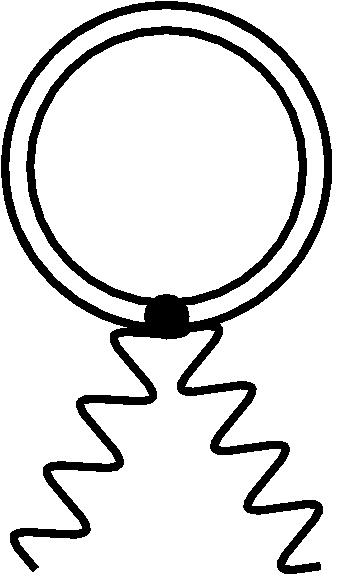}\quad \includegraphics[width=0.5\linewidth]{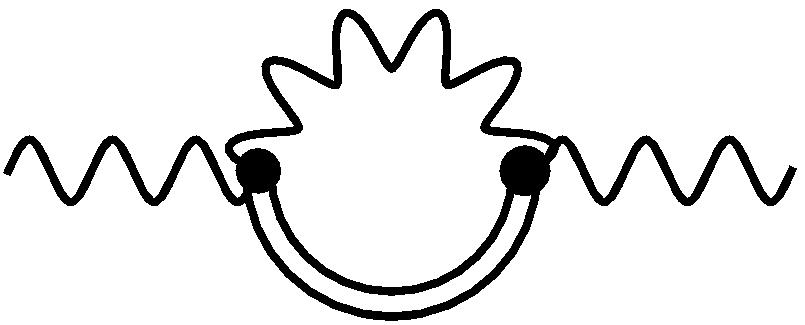}
\end{minipage}
\begin{minipage}{0.45\linewidth}
\caption{\label{fig:QGdiagseta} Double lines denote metric fluctuations and wavy lines gauge bosons. Two diagrams with quantum gravity contributions enter the expression for $\eta_A$.}
\end{minipage}
\end{figure}

In the following, we first consider the TT approximation, which consists of neglecting all modes in $h_{\mu\nu}$ except the transverse traceless mode, for which $\bar{D}^{\nu}h_{\mu\nu}^{\rm TT}=0$ and $h^{\mu\, \, \, \, \rm TT}_{\mu}=0$, cf.~Eq.~\eqref{ttaprox} in App.~\ref{App:vertices}. In that setting for gravity and in the perturbative approximation, it turns out that within our truncation only the quantum-gravity diagrams in Fig.~\ref{fig:QGdiagseta} contribute, yielding
\bea
\beta_{\rho_3}&=& \beta_{\rho_4}= \frac{\eta_A}{2}{\rho}= \frac{1}{48 \pi^2}{\rho^3} - G\, {\rho} \frac{5}{36\pi}\left(\frac{8}{1- 2\Lambda} + \frac{8 - \eta_h}{(1-2\Lambda)^2}\right). 
\eea
For positive $G$, $\Lambda<1/2$  (and $\eta_h<16(1-\Lambda)$ as is also required for our choice of regulator, see \cite{Meibohm:2015twa}), the quantum-gravity contribution renders the gauge coupling asympotically free. For positive $\eta_h$, the critical value of $\Lambda$, for which quantum gravity counteracts asymptotic freedom, is shifted further
away from
the boundary at $\Lambda=1/2$. Evidence for quantum-gravity induced  asymptotic freedom was already found in \cite{Daum:2009dn,Harst:2011zx,Folkerts:2011jz,Christiansen:2017gtg}, where that same conclusion was reached for Abelian and non-Abelian gauge theories.

\begin{figure}[!t]
\begin{minipage}{0.6\linewidth}
\includegraphics[width=0.9\linewidth]{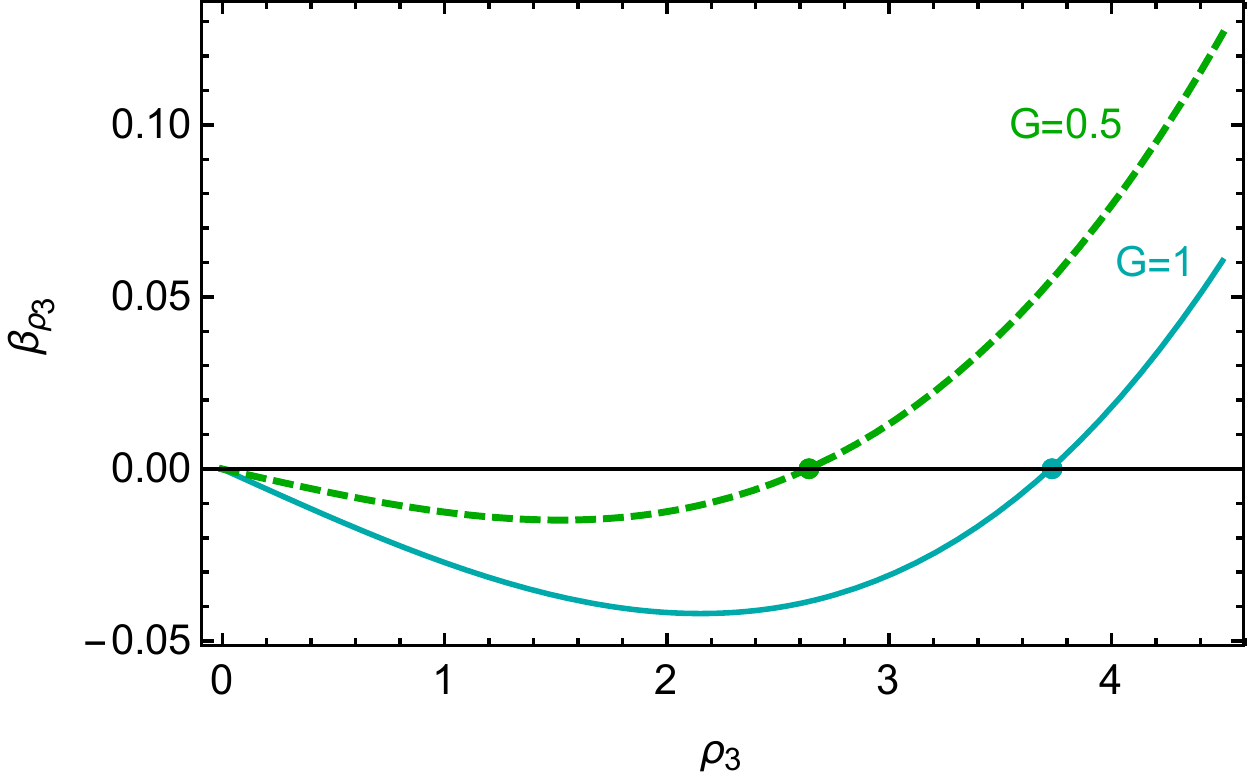}
\end{minipage}
\begin{minipage}{0.38\linewidth}
\caption{\label{fig:betarho3TT} The beta-function for $\rho$ exhibits an asymptotically  safe fixed point under the impact of quantum gravity, cf.~Eq.~\eqref{eq:FPrho3TT} that depends on the microscopic values of the gravity parameters. Here we set $\eta_h=0$, $\Lambda=-6$.}
\end{minipage}
\end{figure}

Moreover, the interplay of quantum gravity and matter degrees of freedom results in an asymptotically safe fixed point at 
\bea
\rho^{\ast}= \frac{\sqrt{\frac{20\pi}{3} G}\sqrt{16(1- \Lambda)- \eta_h}}{1-2\Lambda}.\label{eq:FPrho3TT}
\eea

As the free fixed point is UV attractive in $\rho$, the interacting fixed point must be UV repulsive. A structure with two fixed points, one of them UV repulsive and the other UV attractive, necessarily leads to constraints on the IR values of the coupling, as the basin of attraction of the UV attractive fixed point is bounded by the UV repulsive fixed point. In our case, this leads to an \emph{upper} bound on the IR value of the coupling. This becomes clear from the form of the beta-function, cf.~Fig.~\ref{fig:betarho3TT}: For IR values of the coupling above $\rho^{\ast}$, the flow drives the coupling to ever increasing values, triggering the approach to Landau-pole like behavior, which indicates a breakdown of the theory. Conversely, all IR values below $\rho^{\ast}$ lead to asymptotically free behavior in the IR. Finally, for $\rho=\rho^{\ast}$ in the IR, the flow reaches the interacting fixed point in the UV. In the more realistic case, where gravity is "switched on" dynamically in the vicinity of the Planck scale, the critical IR value $\rho_{\rm crit}$ is shifted away from $\rho^{\ast}$, as we will see in Sec.~\ref{sec:flows}, where we include the additional degrees of freedom of the Standard Model and consider the running of $G, \Lambda$.

\subsection{Quantum-gravity contribution to the running gauge coupling}

In our setup of the flow equation, the gravity contribution to the beta function for the gauge coupling depends on the gauge parameters $\alpha, \beta$ and $\xi$. $\alpha=0$ and $\xi=0$ are preferred choices, as they correspond to hard implementations of the gauge condition and thus to fixed points of the RG flow \cite{Litim:1998qi, Ellwanger:1995qf, Lauscher:2001ya}. The gauge parameter $\beta$ rotates the contribution of scalar fluctuations of the metric between the two scalar modes. We constrain admissable choices by the requirement that $\beta_{\rho_3} = \beta_{\rho_4} = \frac{\eta_A}{2}\, \rho$, which uniquely selects $\beta=1$ while simultaneously resulting in a $\xi$-independent form of the beta function. Incidentally, in \cite{Gies:2015tca} this choice was shown to be close to an extremum in the critical exponents in the pure gravity fixed point in the Einstein-Hilbert truncation, indicating a preference of this value according to the principle of minimum sensitivity. For a study of the full gauge dependence, see App.~\ref{sec:QG3point}.

\begin{figure}[!t]
\begin{minipage}{0.6\linewidth}
\includegraphics[width=0.9\linewidth]{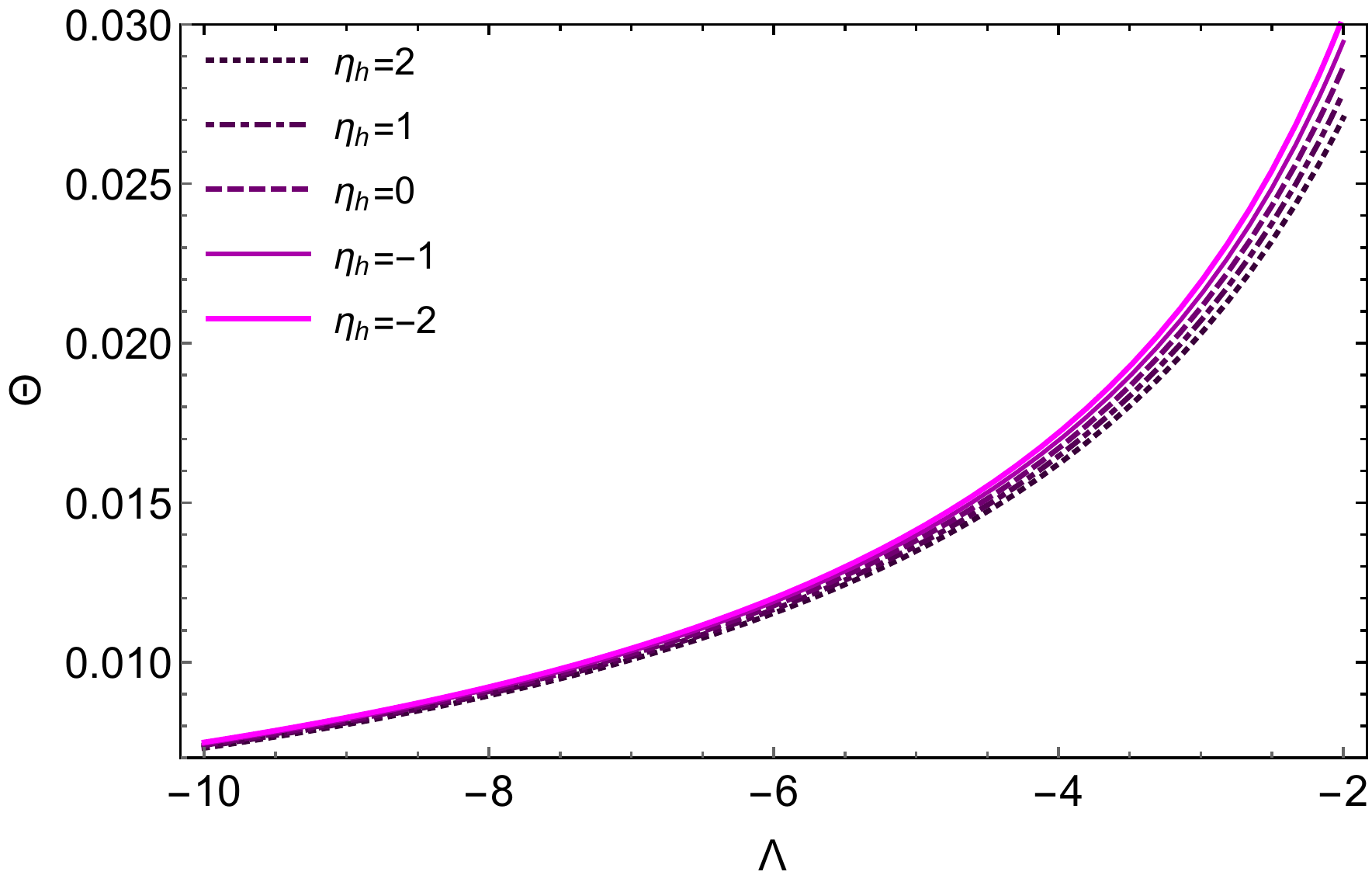}
\end{minipage}
\begin{minipage}{0.38\linewidth}
\caption{\label{fig:theta34_lambda_etah_tail} Significantly different choices of $\eta_h$, parameterizing effects beyond our truncation, only lead to slight quantitative changes of the critical exponent $\Theta$ in the regime $\Lambda<-1$ that we focus on in Sec.~\ref{sec:flows}.}
\end{minipage}
\end{figure}

Using that gauge and the perturbative approximation for the matter fields, such that all $\eta_{A/\phi}$ from scalar-derivatives of the regulator are set to zero, we obtain
\bea
\label{eq:beta_pref}
\beta_{\rho_3}\Big|_{\rm grav} = \frac{1}{48\pi^2}\, \rho^3 - \frac{G\, (4(1-4\Lambda)+\eta_h)}{16\pi(1-2\Lambda)^2}\, \rho.
\eea
 The quantum-gravity contribution is exactly the one obtained in \cite{Christiansen:2017gtg}, if the additional coupling $w_2$ in that work is neglected. As in the TT approximation, an IR attractive fixed point at $\rho^{\ast}>0$ is induced if $\Lambda<(4+\eta_h)/16$. The critical exponent has the opposite sign, but same magnitude as the one at the free fixed point,
 \be
 \Theta_{\rm int}= - \Theta_{\rm free} :=\Theta= - \frac{\partial \beta_{\rho}}{\partial \rho}\Big|_{\rho=\rho^{\ast}}= -G\frac{4-16\Lambda + \eta_h}{8\pi(1-2\Lambda)^2}.
 \ee
 In the next section, we focus on the regime $\Lambda<<0$. We observe that in this regime our results appear to be quite robust under extensions of the truncation which lead to changes of $\eta_h$, cf.~Fig.~\ref{fig:theta34_lambda_etah_tail}. Physically, this regime is one where quantum gravity fluctuations are present but not strong in the sense that the effective gravitational coupling $\frac{G\, (4(1-4\Lambda)+\eta_h)}{16\pi(1-2\Lambda)^2}$ that enters the beta function for the gauge coupling, is suppressed at negative $\Lambda$. This suppression accounts for the robustness of our result with respect to changes in $\eta$.

Let us briefly comment on the relation of our result to that of the EFT setting \cite{EFTabelian}, where one-loop gravity corrections might be removable by a field redefinition \cite{Ellis:2010rw}. However, at an interacting, asymptotically safe fixed point, the use of field redefinitions to identify inessential couplings is a little more involved: a general field-redefinition mixes eigendirections at the fixed point. Thus only those couplings, that are at the same time eigendirections at the fixed point and can be removed by a field redefinition, are in fact inessential. As emphasized, e.g., in \cite{Rosten:2010vm}, operators which are redundant at one fixed point, and removable by a field redefition, are not necessarily redundant at another fixed point. Moreover, a nonzero gravity contribution to the physical running at one loop remains in the EFT case in the presence of a cosmological constant \cite{Toms:2009vd} which is an essential ingredient of our setting.

\section{Upper bound on the U(1) hypercharge coupling in the Standard Model}\label{sec:flows}
 We now focus on the Abelian gauge coupling in the Standard Model, the U(1) hypercharge coupling $g_Y$. Schematically,
\be
\beta_{g_Y} = \frac{g_Y^3}{16\pi^2} \frac{41}{6}-g_Y\, f_g,
\ee
where $f_g$ is a parameterization of the quantum-gravity contribution that we have evaluated in a specific truncation in the previous section.

Then, the fixed-point value for an interacting fixed point is given by 
\be
\label{eq:fp_gy}
g_{Y}^{\ast}= 4 \sqrt{\frac{6}{41} f_g}\pi
\ee
Starting from that fixed point in the UV, quantum gravity fluctuations force the coupling to remain at its fixed-point value all the way down to the Planck scale. At the Planck scale, quantum gravity fluctuations switch off rapidly, and the usual Standard Model running of the coupling takes over, resulting in a unique IR value of the gauge coupling, which we read off at the top mass scale, i.e.,
\be
g_{Y\, \rm IR} = g_Y(k=173\, \rm GeV).
\ee
A second, free and UV attractive fixed point exists for $f_g>0$.

The interacting fixed point has a two-fold role in our scenario: On the one hand, it provides a prediction of the IR value of the coupling along one unique trajectory, emanating from a fully interacting fixed point. On the other hand, it results in a strict upper limit on the IR values of the gauge coupling which can be reached from a UV complete model. The second property is a consequence of the interplay between the free and the interacting fixed point:
The free fixed point is UV attractive. Thus, it can be connected to a range of IR values of the coupling. That range is bounded from above by the unique critical IR value $g_{Y,\,\rm crit }$ of the coupling that is connected to the interacting fixed point in Eq.~\eqref{eq:fp_gy}. As this fixed point is UV repulsive, it shields IR values above the critical value $g_{Y,\,\rm crit}$, such that they cannot be connected to the free fixed point. Thus, IR values $g_{Y\, IR}>g_{Y,\,\rm crit}$ cannot be reached from any of the two UV fixed points, i.e., they correspond to models where new physics must exist at higher scales. Thus, the interacting fixed point provides an upper bound on the IR values of the gauge coupling in a UV complete model, cf.~Fig.~\ref{fig:upperboundgy}. Moreover, using the universality class defined by the interacting fixed point leads to a UV complete model with a reduced number of free parameters, as the IR value of the coupling is fixed uniquely.

The requirement that the predicted IR value of the coupling matches the experimental result translates into the constraint
\be
f_g= \frac{0.096}{\pi^2}.\label{eq:fgval}
\ee
The constraint Eq.~\eqref{eq:fgval} on the gravitational parameter space arises if one demands that asymptotically safe gravity should provide a prediction of the U(1) coupling. 
Observational viability demands that $f_g \geq \frac{0.096}{\pi^2}$ \emph{must} be satisfied, as otherwise the experimentally observed value of the hypercharge coupling cannot be connected to any of the two fixed points in the UV. In the future, it is crucial to analyze whether extended truncations with the general form of the metric propagator and the impact of Standard Model matter fluctuations do in fact converge to fixed-point values such that $f_g\gtrsim 0.096/\pi^2$ is satisfied.

Physically, the effect of quantum fluctuations of gravity which result in a UV completion for the gauge system can be understood as a form of effective dimensional reduction: The quantum-gravity term is \emph{linear}  in the gauge coupling, just as a term that arises from a nonvanishing scaling dimension. $f_g>0$ acts like a scaling dimension in $d<4$ dimensions. In our approximation, a UV complete model can only be obtained if the effective scaling dimension is that of a dimensionally reduced gauge theory. We conjecture that this dimensional reduction of the matter system away from its critical dimensionality $d=4$ towards the critical dimensionality of gravity, $d=2$, is crucial for a viable UV completion. \\
It is tempting to speculate that there might be a connection to dimensional reduction in the spectral dimension to $d_s=2$ as has been observed in asymptotic safety \cite{Lauscher:2005qz,Reuter:2011ah,Calcagni:2013vsa} and is derived from UV scale invariance. Moreover, dimensional reduction appears as a potentially universal result from different quantum-gravity approaches \cite{Carlip:2016qrb,Ambjorn:2005db,Lauscher:2005qz,Horava:2009if,Reuter:2011ah,Calcagni:2013vsa,Carlip:2015mra}. Here, we find hints of a different form of effective dimensional reduction in  asymptotically safe gravity, which is linked to the effective scaling dimension of matter couplings.

We have identified the gauge $\beta=1, \alpha=0$ as a preferred choice of gauge, as it respects $\beta_{\rho_3} = \beta_{\rho_4} = \frac{\eta_A}{2}\rho$ also for the gravity contribution, cf.~App.~\ref{sec:QG3point}. We  now make the assumption -- that should be tested in future work -- that this will still remain the case once we analyze the running of the gauge coupling including charged fermions. In that case, there will be a photon-fermion-antifermion vertex as well as a two-photon-fermion-antifermion vertex which can be used to read off the running coupling. Here, we add the contribution of charged fermions with the quantum numbers that they have in the Standard Model, so that $\rho= g_Y$ will actually be the hypercharge coupling.

Then we obtain 
\be
\beta_{g_Y} = \frac{g_Y^3}{16\pi^2} \frac{41}{6}- g_Y\,G \frac{1-4\Lambda}{4\pi(1-2\Lambda)^2},\label{eq:betagY}
\ee
{ i.e., in our truncation, 
\be
f_g = G \frac{1-4\Lambda}{4\pi(1-2\Lambda)^2},
\ee
holds, but of course the dependence of $f_g$ on $\Lambda$ will change, once higher-order effects are included.  We stress that we parameterize higher-order contributions in the metric propagator by $\Lambda$, as the fixed-point value for the U(1) coupling only depends on the full propagator, but not on $G, \Lambda$ etc.~\emph{separately}. In fact, the two diagrams in Fig.~\ref{fig:QGdiagseta} simply depend on the integrated metric propagator, and we can summarize their effect.
This yields an interacting, UV repulsive and therefore predictive fixed point at
\be
g_Y^{\ast}=\sqrt{\frac{24\pi}{41}}\sqrt{\frac{G(1-4\Lambda)}{(1-2\Lambda)^2}}.
\label{eq:gyast}
\ee
 
As the fixed-point value of $g_Y$ depends on the UV value of the gravity couplings, each point in the $G, \Lambda$ plane is mapped onto an IR value of $g_Y$ along the corresponding critical trajectory, cf.~Fig.~\ref{fig:glambdaplane}.  Since the fixed-point value $g_{Y}^{\ast}$ in Eq.~\eqref{eq:gyast} depends on the gravity parameters, the value of the upper bound on the IR coupling becomes sensitive to the \emph{microscopic} gravity parameters. Under the assumption of no new physics on intermediate scales, this provides a link between IR observables and the microscopic dynamics of spacetime. The often-heard assertion that it is impossible to constrain quantum gravity observationally, typically based on a comparison of experimentally accessible scales and the Planck scale, thus appears to be too pessimistic. A huge gap of scales can be bridged by the Renormalization Group flow, which links UV physics and IR physics. In particular, it allows us to map possible microscopic models onto their macroscopic counterparts. Conversely, experimental results at low (compared to the Planck scale) energies can therefore constrain the microscopic properties of quantum gravity.

\begin{figure}[!t]
\begin{center}
\includegraphics[width=0.6\linewidth]{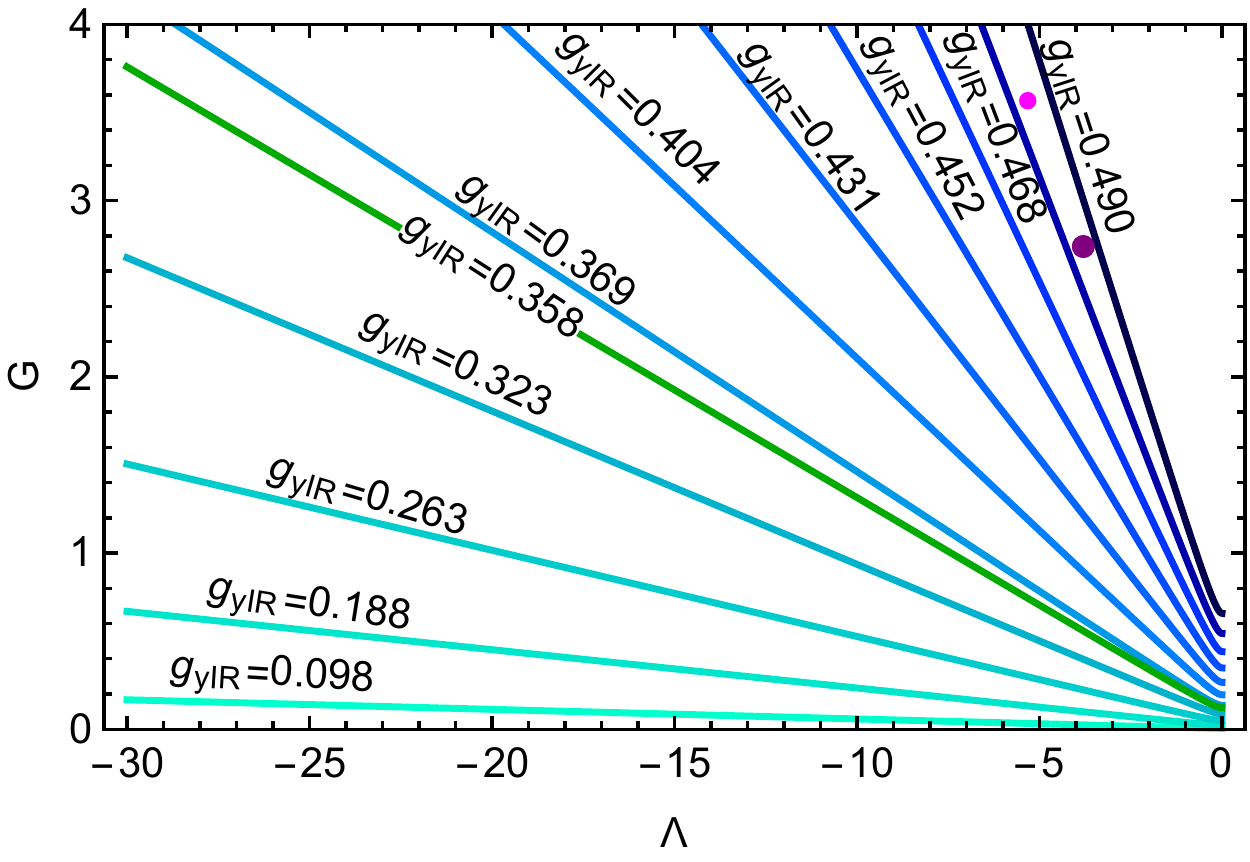}
 \end{center}
\caption{\label{fig:glambdaplane} Lines of constant IR values for $g_Y$. The value extracted from measurements is highlighted in green, at $g_Y=0.358$. The larger purple dot is the fixed point obtained from Eq.~\eqref{eq:gravityFP} with Standard Model matter content, whereas the smaller magenta dot includes the effect of three additional Weyl fermions that are required to accommodate neutrino masses. To obtain the IR values, we use that $g_Y(k=M_{\rm Pl})=g_Y^{\ast}$ and use the one-loop beta function Eq.~\eqref{eq:betagY} with $G=0$ to integrate the flow from the Planck scale to the IR scale. The structural similarity to the dependence of the IR value of the top mass on the microscopic gravity parameters \cite{Eichhorn:2017ylw} is due to a similar form of the beta functions, with the only difference being the numerical prefactor of the gravity contribution as well as the critical value of $\Lambda$, at which the critical exponent switches its sign and an interacting, predictive fixed point is induced. }
\end{figure}

\begin{figure}[!t]
\begin{center}
\includegraphics[width=0.6\linewidth]{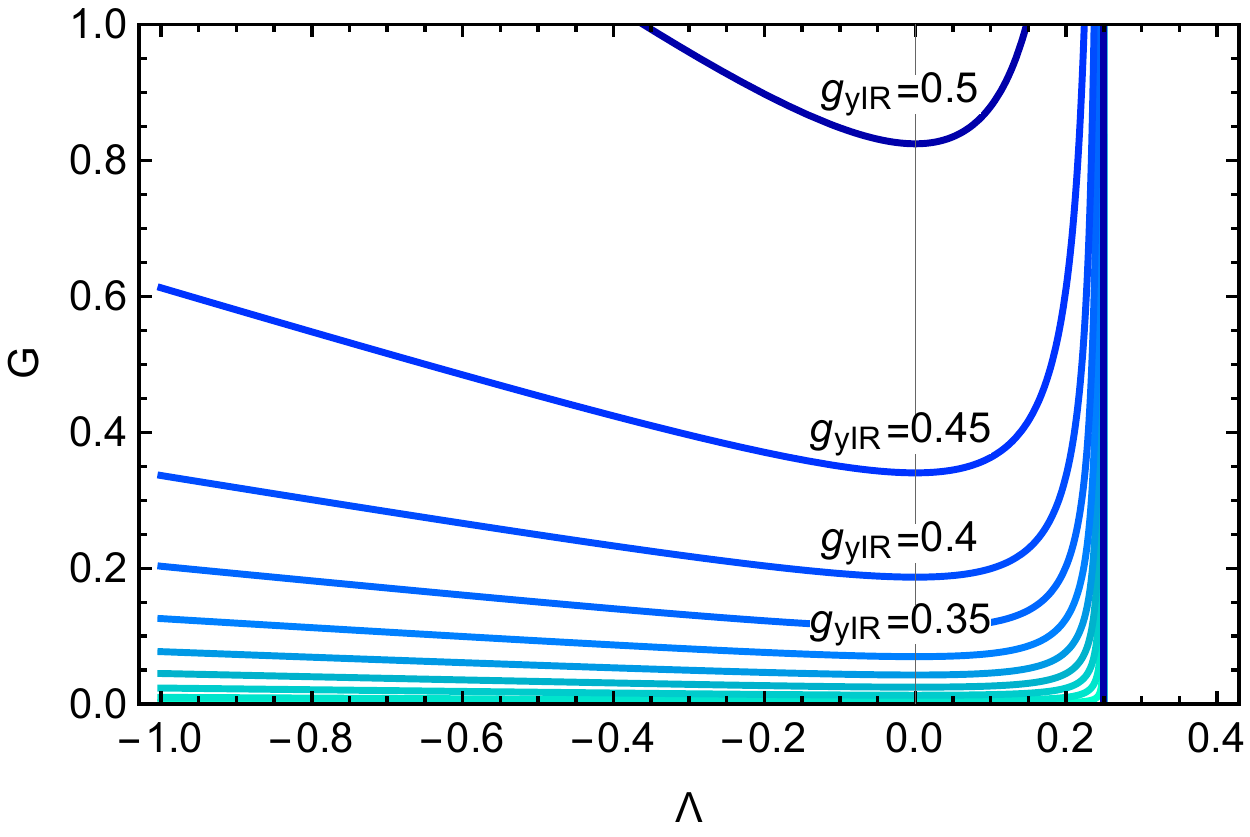}
 \end{center}
\caption{\label{fig:glambdaplane_pos} Lines of constant IR values of $g_Y$ in a zoom in the region around $\Lambda \approx 0$ for vanishing anomalous dimension of the metric $\eta_h=0$. Within our truncation, lines of different IR values of $g_Y$ are pulled closer together as $\Lambda$ approaches the critical value at which the interacting fixed point merges with the free one. This moves the phenomenologically preferred values of $G$ to significantly smaller values.}
\end{figure}

The second, free fixed point exists for all values of $G$ and all values of $\Lambda$ except those too close to the pole at $\Lambda=1/2$. Within our truncation, $\Lambda=1/4$ is the critical value at which the gravity contribution switches sign, as already observed in \cite{Folkerts:2011jz}.

For our quantitative analysis, we are particularly interested in the region $\Lambda<0$, where a prediction of the top mass and Higgs mass from asymptotic safety might be possible \cite{Shaposhnikov:2009pv, Bezrukov:2012sa,Eichhorn:2017ylw}. It is a highly nontrivial result of our analysis, that the same region of the gravity parameter space in our truncation also allows a prediction of the IR value of the U(1) coupling which is in the vicinity of the experimentally observed value, cf.~Fig.~\ref{fig:glambdaplane}.

 It is an open question whether extensions of the truncation will provide fixed-point values in this regime and whether higher-order effects on the gauge coupling will remain small. However, we observe that within this regime the IR value of $g_Y$ is not strongly sensitive to the microscopic values of the gravity parameters. Therefore, if the fixed-point values for gravity converge in the regime $\Lambda<0$, the prediction for $g_Y$ in the IR varies only slightly under variations in $G, \Lambda$. This signals a robustness of our result under variations of the truncation.

The weakly-coupled gravity regime required by a quantitatively accurate prediction of the IR value of the hypercharge coupling, as well as the top mass \cite{Eichhorn:2017ylw} within our truncation, is simultaneously a regime where first extensions of the truncation show signs of stability: As gravity is weakly-coupled, all effects of induced couplings appear suppressed. 
Specifically, the effect of induced photon self-interactions on $g_Y$ appears to be negligible, as the inducing quantum-gravity effects are of course also suppressed at $\Lambda<<0$. In particular, we find that the fixed-point value for photon self-interactions, parameterized by a  $F^4$ interaction term, is $w_2^{\ast} = -0.025$ in the conventions from \cite{Christiansen:2017gtg}. Accordingly, the back-coupling of $w_2$ into the critical exponent and thus fixed-point value of $g_Y$ is suppressed by an order of magnitude in comparison with the direct gravity contribution in Eq.~\eqref{eq:fgval}. 

We combine Eq.~\eqref{eq:betagY} with beta-functions for $G, \Lambda$, to obtain the flow of the coupled system. Note that the back-coupling of $g_Y$ into the flow of the gravity couplings only sets in at subleading order  within the background approximation: The main contribution to the running of the gravity couplings comes from minimally coupled matter and gauge fields. The interaction between those, such as that mediated by $g_Y$, only impacts the anomalous dimensions of the matter fields. The anomalous dimensions of matter and gauge fields in turn only appear as a correction in the loop term, which is subdominant compared to the main term. Thus, we assume that the approximation in which only the effect of minimally coupled matter is included in the beta functions for gravity is sufficient for our purposes. We use beta functions in a single-metric approximation with $\alpha=0, \beta=1$ and vanishing anomalous dimension $\eta_h=0$ from \cite{Lippoldt} with the matter contribution from \cite{Dona:2013qba}, which read
\bea
\beta_G &=& 2G + \frac{G^2}{6\pi} \left(2N_D +N_S - 4N_V \right) - \frac{G^2}{6\pi} \left(14 + \frac{6}{1-2\Lambda}  + \frac{9}{(1-2\Lambda)^2}\right),\\
\beta_{\Lambda} &=& -2 \Lambda +\frac{G}{4\pi}\left(N_S-4 N_D + 2 N_V \right) + \frac{G}{6\pi}\Lambda \left(2N_D+N_S-4N_V \right)\nonumber\\
&{}& - \frac{3}{2\pi}G - \frac{7}{3\pi}G\, \Lambda- \frac{3 G}{4\pi (1-2\Lambda)^2}+ \frac{7G}{4\pi(1-2\Lambda)}.
\eea

As a main difference to the analysis in \cite{Harst:2011zx}, we include the effect of all matter degrees of freedom on the gravitational fixed point value. Within the single-metric approximation, the inclusion of fermions results in a shift of the fixed point towards large negative $\Lambda$ \cite{Dona:2012am,Dona:2013qba,Eichhorn:2016vvy,Biemans:2017zca}. In turn, this leads to a signficantly smaller value of $f_g$, shifting the fixed-point value of $g_Y$ and accordingly also its IR value much closer to the observed value.

For the Standard Model, where $N_S=4, N_D=45/2, N_V=12$, we obtain
\be
G^{\ast}=2.73, \, \Lambda^{\ast}=-3.76, \, g_Y^{\ast}=1.05.\label{eq:gravityFP}
\ee
This results in a flow which leads to
\be
g_{Y\, \rm IR}= 0.487,
\ee
as the prediction on the asymptotically safe trajectory, that simultaneously serves as the upper bound on viable IR values of the coupling. Our result lies within $35 \%$ of the experimentally observed value, which we consider reachable by extensions of the truncation from our simple leading-order approach. Most importantly, it suggest that asymptotic safety is compatible with the actual value of the Abelian gauge coupling in the Standard Model that is inferred from measurements.

\section{Conclusions and outlook}\label{sec:outlook}
Within a truncation of the RG flow, we provide further evidence for a scenario in which a \emph{weakly-coupled} UV completion for quantum gravity and matter might exist beyond the Planck scale. The microscopic model is defined at an interacting fixed point with scaling properties close to canonical scaling. In particular, in the corresponding regime of microscopic gravitational couplings, quantum gravity induces a partially interacting fixed point for the Standard Model couplings with an enhanced predictive power. Specifically, in this work we discover that there is a regime in which quantum gravity generically induces a \emph{perturbative} interacting fixed point for the U(1) hypercharge coupling. Such a fixed point can be matched onto the finite, perturbative value of the U(1) hypercharge coupling that SM running predicts at the Planck scale.
For a large range of microscopic gravity couplings, the resulting, predicted IR value of the gauge coupling is close to the value inferred from measurements.
\\
It is intriguing that the regime of microscopic gravity couplings for which the uniquely determined IR value of $g_Y$ that results from the asymptotically safe fixed point is close to the observed one, is also the regime in which asymptotically safe quantum gravity triggers a predictive fixed point in the top Yukawa coupling and quartic Higgs coupling with IR values of the masses rather close to the observed ones \cite{Eichhorn:2017ylw}. Together, this highly nontrivial fixed-point structure implies that a fascinating scenario could becomes feasible, in which quantum gravity together with the Standard Model is UV complete and the underlying asymptotically safe fixed point has a high predictive power, reducing the number of free parameters of the Standard Model. 
A joint fixed point in the gravity couplings, the top Yukawa and the hypercharge coupling might even trigger a predictive fixed point in the bottom Yukawa which predicts the bottom mass \cite{Held2017}. If such a scenario can be confirmed in extended truncations of the RG flow, it does of course not correspond to a ``smoking-gun signal" of asymptotically safe quantum gravity. On the other hand, a prediction of the seemingly unrelated values of U(1) hypercharge, Higgs mass, top mass and bottom mass from an asymptotically safe fixed point is at least a highly surprising coincidence and definitely deserves detailed further studies.\newline\\

\noindent\emph{Acknowledgements}\\
We thank A.~Held and J.~M.~Pawlowski for insightful discussions.
This research is supported through the Emmy Noether program of the DFG under grant no.~Ei-1037/1. F.~V.~acknowledges support by the HGSFP. 

\begin{appendix}

\section{Beta functions for general gauge parameters}\label{app:betas}

The full beta functions as derived from the 3-vertex and 4-vertex are found to be

\bea
\beta_{\rho_3}&=& -\frac{34 + 3\eta_A + 3\eta_\phi +\xi(12 - 4\eta_\phi)}{96\pi^2}\, \rho_3^3 + (3+\xi)\, \frac{12 - \eta_\phi-\eta_A}{96\pi^2}\, \rho_3\, \rho_4^2\\
&{}& + \frac{G}{96\pi\, (1-2\Lambda)}\, \left[2A\, (8-\eta_\phi) +B\, (8-\eta_A) + \frac{2A\, (8-\eta_h)+B\, (4-\eta_h)}{1-2\Lambda} \right]\, \rho_3,\nonumber\\
\beta_{\rho_4}&=& (3+\xi)\, \frac{12 - \eta_\phi-\eta_A}{192\pi^2}\, \rho_4^3 -\frac{1}{192\pi^2}(\xi\, (24-\frac{\eta_A}{2}-5\eta_\phi) + 32  + 3\eta_A + 3\eta_\phi)\, \rho_3^2\, \rho_4 \nonumber\\
&{}&+ \xi\, \frac{40-\eta_A - 3\eta_\phi}{640\pi^2}\, \frac{\rho_3^4}{\rho_4}\nonumber\\
&{}&+ \frac{G}{96\pi(1-2\Lambda)}\, \left[A\, (8-\eta_\phi) - B\, (8-\eta_A) + \frac{A\, (8-\eta_h)+B\, (4 - \eta_h)}{1-2\Lambda} \right]\, \rho_4,
\eea
where $A$ and $B$ depend on $\Lambda$ and the graviton gauge parameters $\alpha,\, \beta$ in the following way

\bea
A &=& \frac{2\, \left(32 \alpha^2\,  \Lambda  (1-2 \Lambda )^2\right)}{(2 \alpha\,  \Lambda -1)\,  \left(16 \alpha\,  \Lambda ^2-4 (2 \alpha+3)\,  \Lambda +\beta^2\,  (4 \Lambda +1)-6 \beta+9\right)}\nonumber\\
&{}&\frac{2\, \left(\alpha\,  \left(\beta^2\,  \left(32 \Lambda ^2-12 \Lambda -3\right)-2 \beta\,  \left(8 \Lambda ^2+12 \Lambda -9\right)-80 \Lambda ^2+100 \Lambda -31\right)\right)}{(2 \alpha\,  \Lambda -1)\,  \left(16 \alpha\,  \Lambda ^2-4 (2 \alpha+3)\,  \Lambda +\beta^2\,  (4 \Lambda +1)-6 \beta+9\right)},\nonumber\\
&{}&+\frac{2(\beta-1)^2\,  (-(4 \Lambda -3))}{(2 \alpha\,  \Lambda -1)\,  \left(16 \alpha\,  \Lambda ^2-4 (2 \alpha+3)\,  \Lambda +\beta^2\,  (4 \Lambda +1)-6 \beta+9\right)}\nonumber\\
B &=& 12-\frac{4 (\alpha\,  (4 \Lambda -2)+(\beta-2)\,  \beta-4 \Lambda +3)}{4 \Lambda\,   \left(-2 \alpha+\beta^2-3\right)+16 \alpha\,  \Lambda ^2+(\beta-3)^2}-\frac{4 (1 - \alpha)}{1 - 2 \alpha\,  \Lambda}.
\eea

Note that in the gauge $\alpha=0,\, \beta=1$, we find $A=0$ and $B=6$, simplifying the expression for $\beta_{\rho_3}$ to match ~\eqref{eq:beta_pref}.

Gauge independence must of course hold for physical observables. While the gravity-contribution to the beta-function of the gauge coupling is gauge dependent, this does not imply that there is no physical content to this contribution. In fact, it is well-known that beta functions for the Standard Model show a dependence on unphysical choices (gauge, choice of RG scheme) at three loop order and beyond. For dimensionfull couplings, like the Newton coupling, gauge dependence sets in earlier.  In truncations of the FRG, the dependence of presumed observables on unphysical parameters that can arise in truncations can be exploited to test the quality of the truncation, see, e.g., \cite{Gies:2015tca}.

\section{Quantum-gravity contributions to the running gauge coupling from the three-and four-point vertex}\label{sec:QG3point}

\begin{figure}[!t]
\begin{minipage}{0.4\linewidth}
\includegraphics[width=0.9\linewidth,clip=true, trim=28cm 0cm 0cm 18cm]{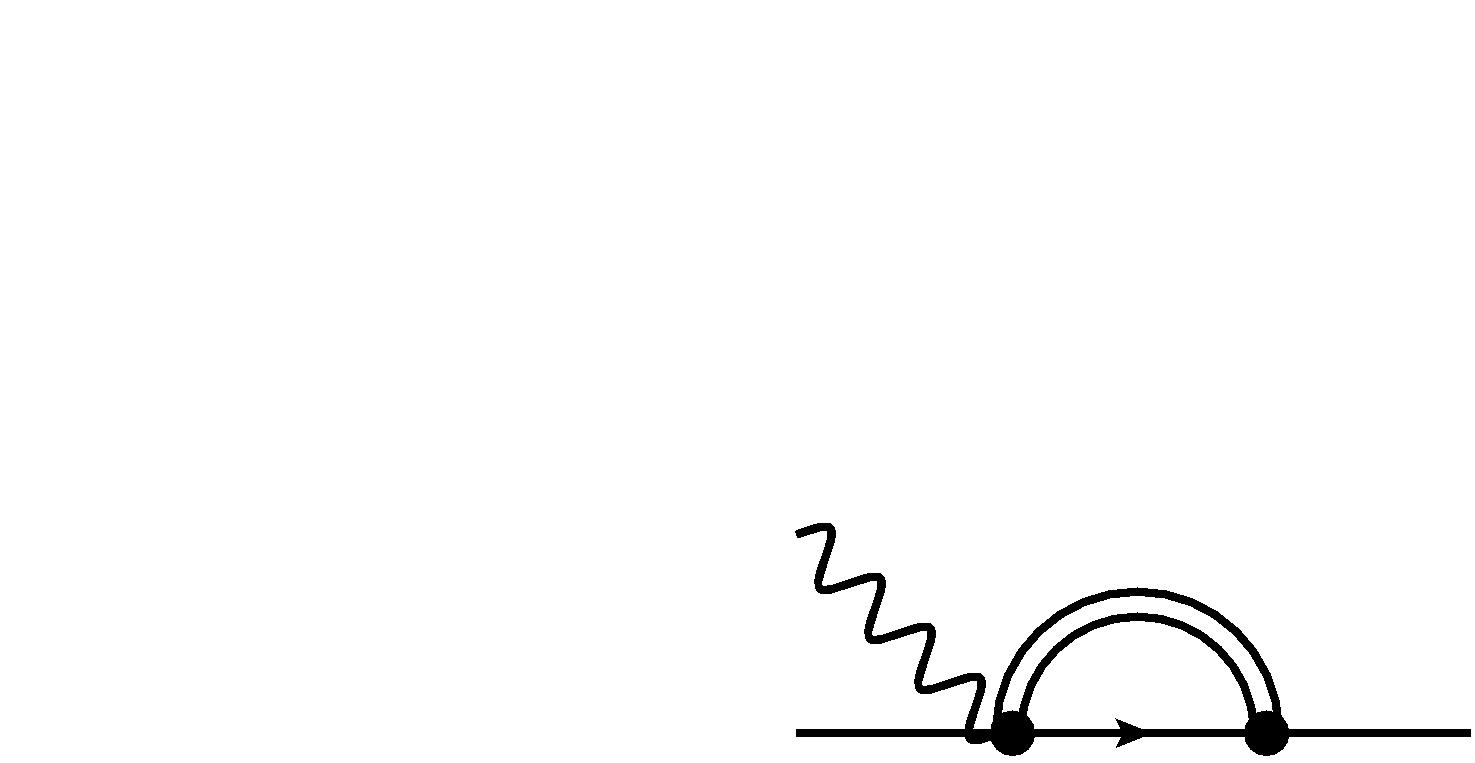}
\end{minipage}
\begin{minipage}{0.55\linewidth}
\caption{\label{fig:betarho3gravdiags} We show the only diagram that contributes to $\beta_{\rho_3}$ and contains metric fluctuations. Other diagrams with internal metric fluctuations and the correct number of external legs vanish when projected onto $\rho_3$ according to Eq.~\eqref{eq:projrho3}.}
\end{minipage}
\end{figure}

\begin{figure}[!t]
\begin{minipage}{0.4\linewidth}
\includegraphics[width=0.9\linewidth,clip=true, trim=25cm 0cm 0cm 19cm]{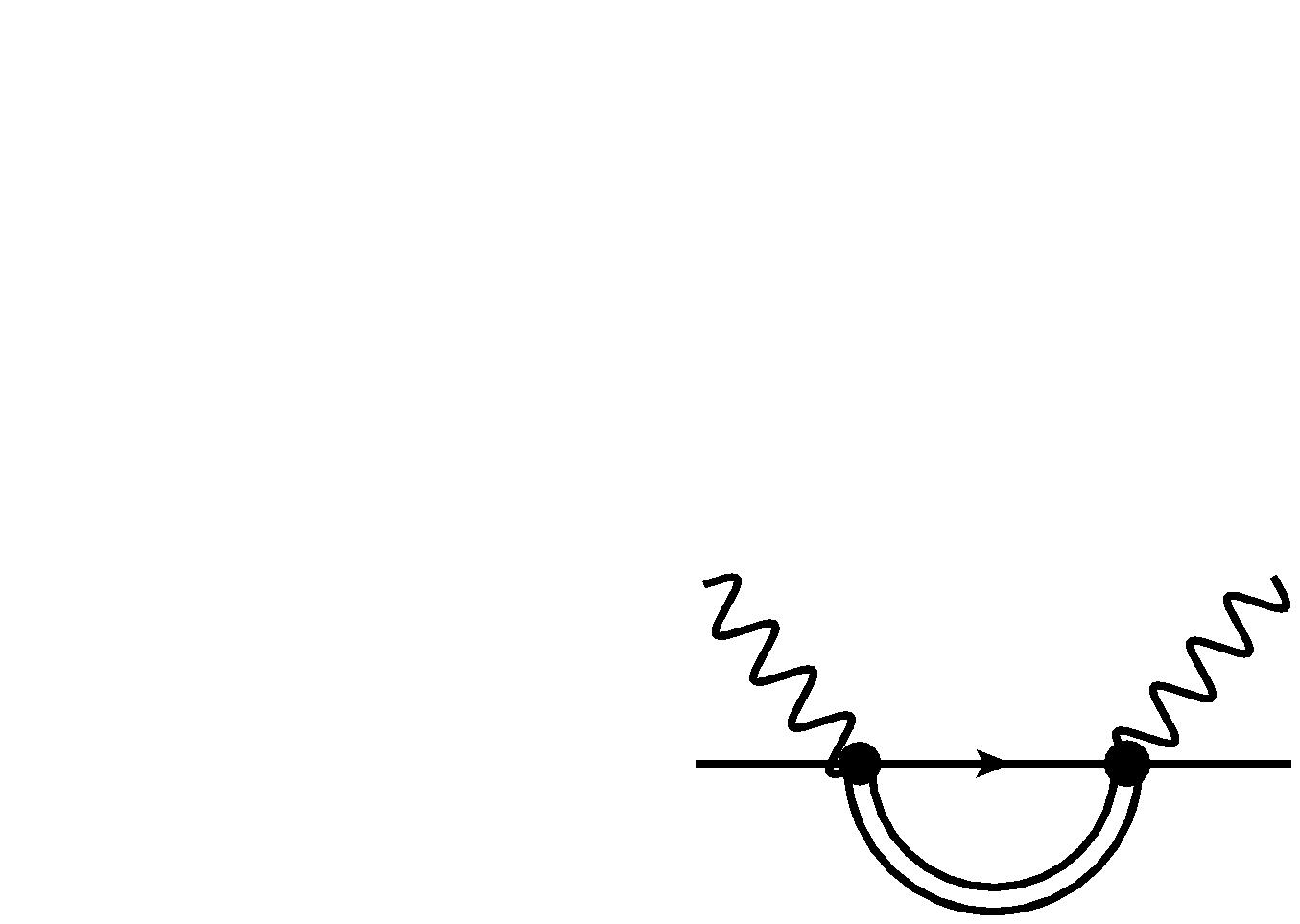}
\end{minipage}
\begin{minipage}{0.55\linewidth}
\caption{\label{fig:betarho4gravdiags} We show the only diagram that contributes to $\beta_{\rho_4}$ and contains metric fluctuations. Other diagrams with internal metric fluctuations and the correct number of external legs vanish when projected onto $\rho_4$ according to Eq.~\eqref{eq:projrho4}.}
\end{minipage}
\end{figure}

 In this section we analyze the gauge dependence of the full beta function.   Even though $\alpha=0$ and $\xi=0$ are preferred as they correspond to hard implementations of the gauge fixing, we will leave them general. 
 Using the expression for $\beta_{\rho_{3/4}}$, given in App.~\ref{app:betas} we investigate the conditions that need to be met in order to satisfy $\beta_{\rho_3} = \beta_{\rho_4} = \eta_A /2\rho$. 
 This provides a  preferred choice for the gauge parameters $\alpha,\, \beta$ that holds for our choice of regulator. This choice confirms the preferred status of the value $\alpha=0$.
 
 We solve $\beta_{\rho_3} = \beta_{\rho_4} = \eta_A /2\rho$ for the gauge parameters $\alpha,\, \beta$, with the full gauge dependent beta functions. For all values of $\Lambda,\, G$ and with the higher order contributions from the anomalous dimensions set to zero, the unique solution to $\beta_{\rho_3} = \beta_{\rho_4} = \eta_A /2\rho$  is given by the gauge choice $\alpha = 0,\, \beta=1$. In \cite{Gies:2015tca} this choice was shown to be close to an extremum in the critical exponents in the pure-gravity fixed point in the Einstein-Hilbert truncation, indicating that it might be a preferred choice of gauge according to the principle of minimum sensitivity.

As a second check, we distinguish $\rho_3$ and $\rho_4$, and analyze their respective fixed points arising from the interplay between gravity and matter fluctuations. In particular, we analyze the critical exponents
at the free fixed point for a range of gauge choices. Defining the critical exponents as
\be
\Theta_{3,4} \equiv -\frac{\partial\, \beta_{\rho_{3,4}}}{\partial\, \rho}\Big|_{\rho = \rho_{3,4}^\ast},
\ee
the behavior of $\Theta_{3,4}$ at the asymptotically free fixed point, $\rho_{3,4}^{\ast}=0$, can be found in Fig.~\ref{fig:theta34_lambda_gauge}, as a function of $\Lambda$. In all gauges both critical exponents tend to zero for large and negative values of $\Lambda$, where the quantum-gravity contribution is more and more suppressed. 
In the asymptotic regime $\Lambda \rightarrow -\infty$, asymptotic freedom is lost.
For $\alpha=\beta=0$, the critical exponents remain positive for all values of $\Lambda$ sufficiently far away from the pole at $\Lambda = 1/2$, but do not coincide. In contrast, the gauge choice $\alpha=0,\, \beta=1$ indicates the exact same behavior for $\Theta_{3},\, \Theta_{4}$  
 which
is independent of $\xi$ for all $\Lambda$, cf.~Fig.~\ref{fig:rho_xi}. The gauge $\alpha=\beta=1$ results in a complete loss of asymptotic freedom for $\beta_{\rho_3}$, whereas $\beta_{\rho_4}$ remains asymptotically free, signalling that this gauge produces inconsistencies in the results. We conclude that together with our choice for the regulator, the gauge $\alpha=0,\, \beta=1$ is preferred.  In addition, we observe a higher degree of universality at sufficiently negative $\Lambda$: While $\Theta_3<0$ for $\alpha=\beta=1$ holds for all $\Lambda$, $\Theta_4$ switches sign and signals asymptotic freedom, in agreement with the critical exponents in other gauge choices.

As a final check, we treat $(\beta_{\rho_3},\, \beta_{\rho_4})$ as a system of equations and look for simultaneous interacting fixed points $(\rho_3^\ast,\, \rho_4^\ast)$. As can be seen from Fig. \ref{fig:rho_xi} , the treatment re-introduces a dependency on the $U(1)$ gauge parameter, $\xi$, at the interacting fixed point. Note that in general $\rho_3^\ast(\xi)\neq\rho_4^\ast(\xi)$ with the exception of the gauge choice $\alpha = 0,\, \beta=1$. 
 
 \begin{figure}[!t]
 \begin{center}
 \includegraphics[width =0.6 \linewidth]{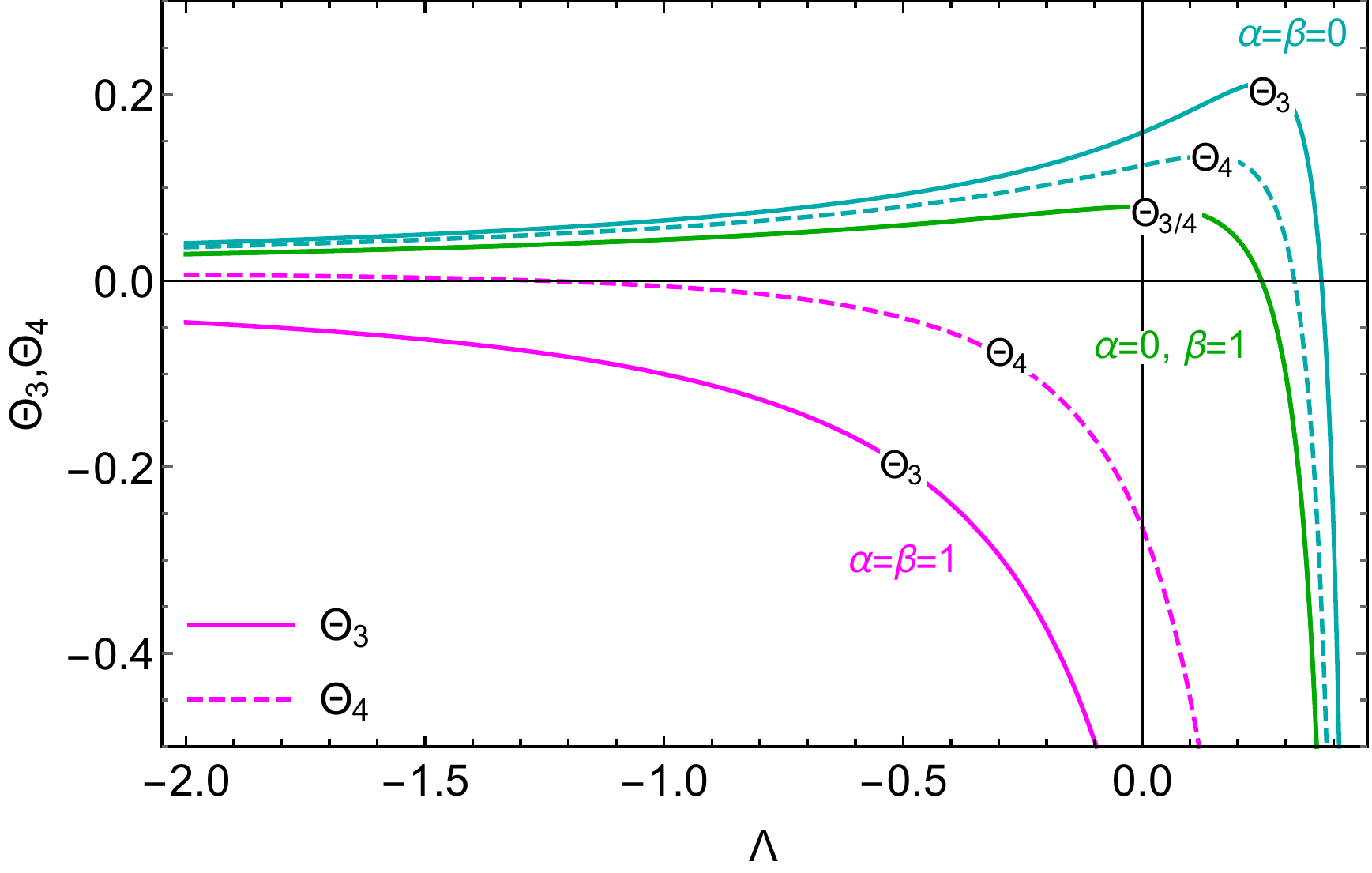}
 \end{center}
 \caption{\label{fig:theta34_lambda_gauge}Critical exponent as a function of $\Lambda$ for different gauges and $G = 1$. The solid and dashed lines correspond to $\Theta_3$ and $\Theta_4$, respectively.}
 \end{figure}

\begin{figure}[!t]
 \begin{center}
\includegraphics[width=0.6\linewidth]{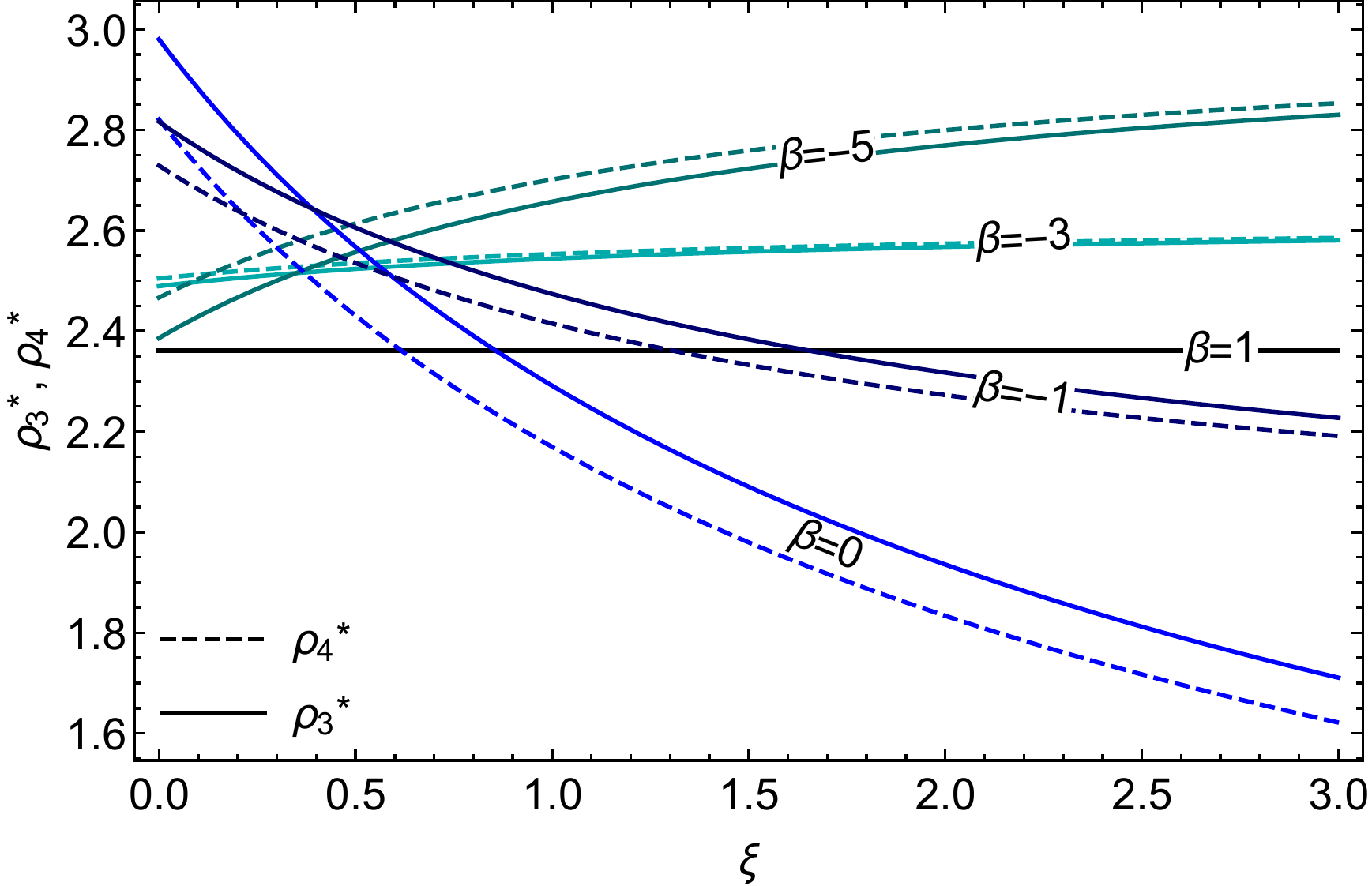}
 \end{center}
\caption{\label{fig:rho_xi}The interacting fixed points, $\rho_{3/4}^\ast$, as a function of the $U(1)$ gauge parameter, $\xi$, for different values of the graviton gauge parameter $\beta$. Here $\Lambda=0,\, G=1,\, \alpha=0$. }
\end{figure}

\section{Beta functions in scalar QED} \label{sec:beta_app}
Here we provide details on how to obtain the beta functions of scalar QED in the FRG framework  and highlight how a nontrivial cancellation of diagrams leads to the expected result of a gauge-independent beta function. Gravitational contributions are added in Sec.~\ref{sec:QGTT} and \ref{sec:QG3point}. 

\begin{figure}[!t]
\begin{center}
\includegraphics[width=0.3\linewidth]{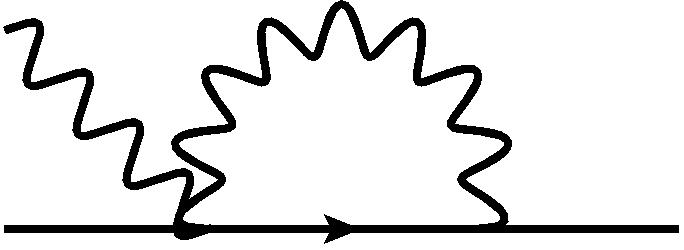}\quad \includegraphics[width=0.3\linewidth,clip=true,trim=28cm 0cm 0cm 25cm]{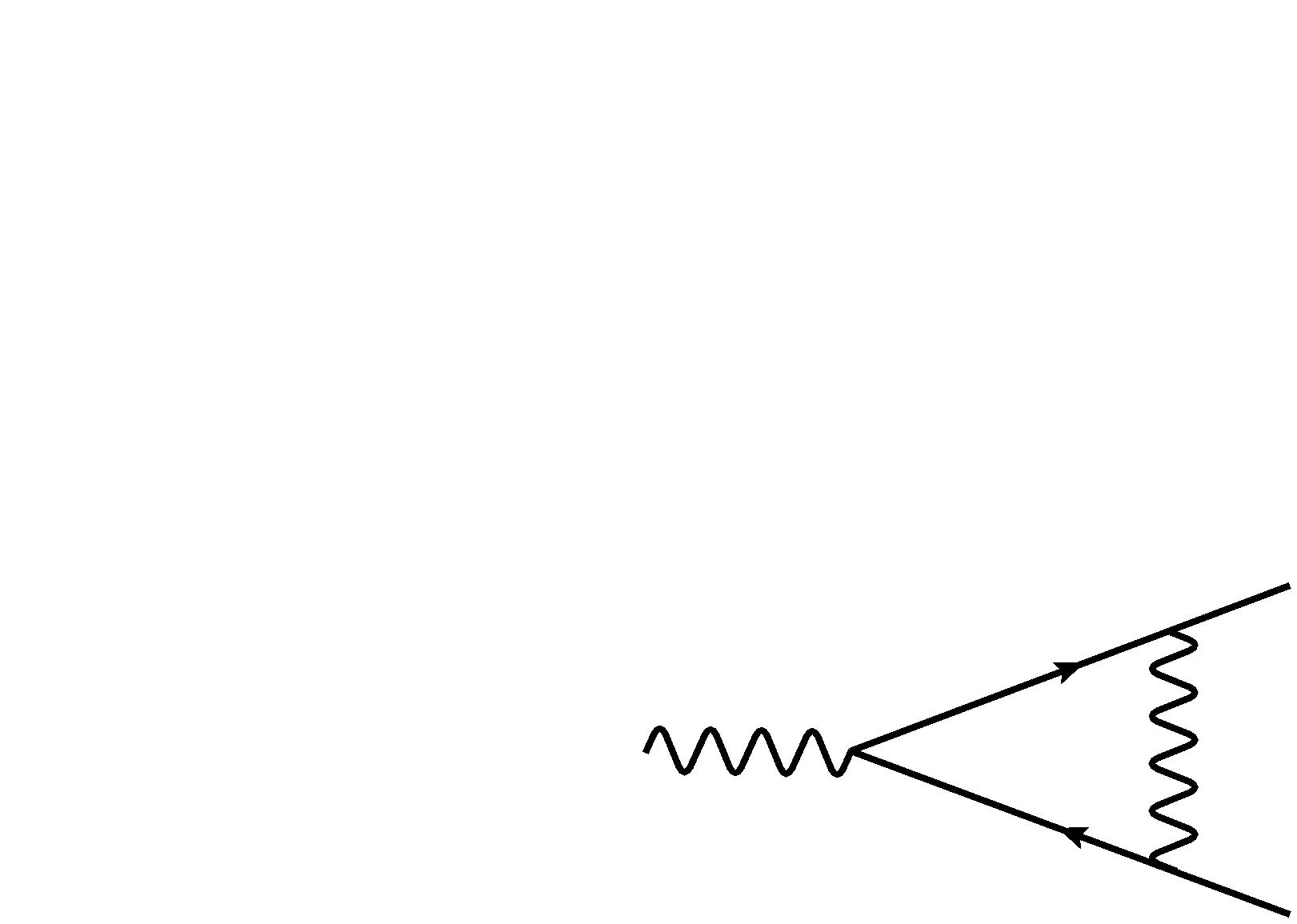}
\end{center}
\caption{\label{fig:betarho3mat} Two diagrams contribute to the running of the three-point vertex. Each diagram is understood to exist in several versions, with $\partial_t R_k$ inserted on each of the internal lines in turn, such that each loop integral is UV finite.}
\end{figure}

Focussing on $\beta_{\rho_3}$, the explicit contributions in Eq.~\eqref{eq:beta_3} arise from 
the two-and three-vertex diagrams in Fig.~\ref{fig:betarho3mat} and can be calculated by using
\bea
\partial_t \rho_3 &= &-\Bigl(\frac{2}{3 q^2}q_1^{\nu}P_{ \nu\sigma}(q_1-q_2)\cdot \frac{\delta^3}{\delta A_{\sigma}(q_1-q_2)\delta\phi^{\dagger}(q_2)\delta\phi(q_1)}\partial_t \Gamma_k\Bigr)\Big|_{q=0, A=\phi^{\dagger}=\phi=0},\label{eq:projrho3}   
\eea
where $|q_i|=q$ and
 we have introduced the transverse projector
\be
P_{\mu\nu}(p) = \delta_{\mu\nu}- \frac{p_{\mu}p_{\nu}}{p^2}.
\ee
For the external momenta, we use a symmetric configuration, see App.~\ref{App:vertices}.

This yields
\bea
\beta_{\rho_3}&=&\rho_{3}\left(\eta_{\phi}+ \frac{\eta_A}{2} \right) +\frac{3 + \xi}{96\pi^2}\left((6 - \eta_A) + (6 - \eta_\phi) \right)\rho_ 3^3 - \frac{\xi}{96\pi^2}\left((8 - \eta_A) + (16 - 2\eta_\phi) \right)\rho_3^3.\nonumber\\
&{}&\label{eq:betarhodirect}
\eea
 The explicit contributions in eq.~\eqref{eq:betarhodirect} are gauge dependent, so we now add the expressions for the anomalous dimensions.

\begin{figure}
\begin{center}
\includegraphics[width=0.3\linewidth]{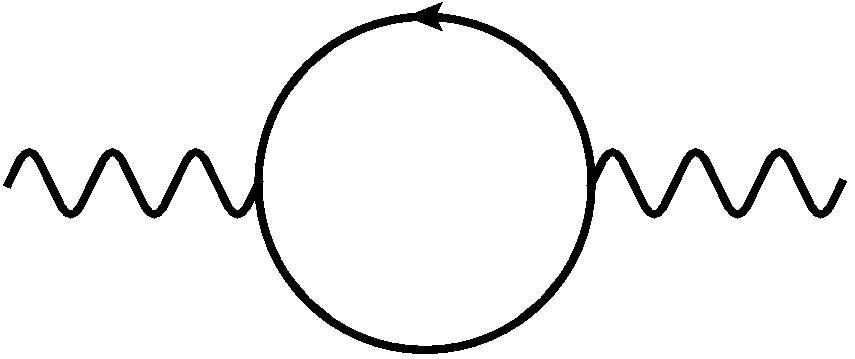}\quad \includegraphics[width=0.3\linewidth]{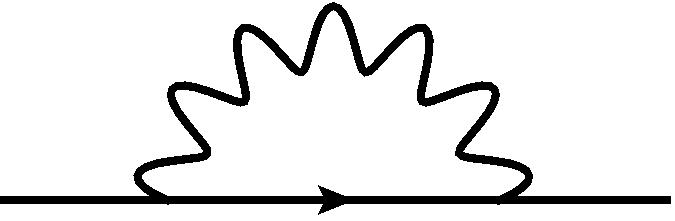}\\
\end{center}
\caption{\label{Fig:etadiags} Diagrams contributing to the flow of the anomalous dimension $\eta_A$ (left) and $\eta_{\phi}$ (right).}
\end{figure} 

To extract the anomalous dimensions, we use that
\bea
\eta_{\phi}&=& - \left(\frac{Z_{\phi}}{p^2} \frac{\delta}{\delta \phi^{\dagger}(p)}\frac{\delta}{\delta\phi(p)}\partial_t\Gamma_k\right)\Big|_{\phi=\phi^{\dagger}= A= h =0, p=0},\\
\eta_A &=&- \left(\frac{P_{\mu\nu}(p)}{Z_A\,3\, p^2}\frac{\delta}{\delta A_{\mu}(p)}\frac{\delta}{\delta A_{\nu}(-p)} \partial_t \Gamma_k\right)\Big|_{A=\phi =\phi^\dagger=h=0,p=0}.
\eea
This projection prescription, applied to the right-hand-side of the Wetterich equation, uniquely extracts the flow of the anomalous dimensions.
Keeping the full dependence on the gauge parameter $\xi$ for the photon, the general result for the anomalous dimensions is 
\bea
\eta_{\phi}& =& - \frac{(3-\xi)}{8 \pi^2}\rho_3^2 -  \frac{\eta_A(3+\xi) + \eta_\phi (3 - 2\xi)}{96\pi^2}\rho_3^2,\label{eq:etaphimat}\\
\eta_A &=& \frac{1}{24\pi^2}\rho_3^2. \label{eq:etaAmat}
\eea
Only $\eta_\phi$ exhibits a gauge dependence. Solving Eq.~\eqref{eq:etaphimat} for $\eta_\phi$ and inserting Eq.~\eqref{eq:etaAmat}, we obtain
\be
\eta_{\phi} =  \frac{\rho_3^2}{24\pi^2}\frac{288\pi^2 \left(-3+\xi\right) - \left(3+\xi \right)\rho_3^2}{(96\pi^2 + \left(3-2\xi \right)\rho_3^2)},
\ee
which includes
terms of higher order in the coupling, starting with $\rho_3^5$. To recover the perturbative one-loop result from the functional RG, higher-order terms in the anomalous dimensions have to be set to zero. Thus, the perturbative result for the anomalous dimensions, which is still gauge dependent, reads
\bea
\eta_{\phi}\Big|_{\rm pert}& =& - \frac{(3-\xi)}{8 \pi^2}\rho_3^2,\\
\eta_A\Big|_{\rm pert} &=& \frac{1}{24\pi^2}\rho_3^2. 
\eea
In the perturbative approximation, where the higher-order terms that arise through the anomalous dimensions, are set to zero, the beta function becomes independent of the $U(1)$ gauge parameter $\xi$.
This entails a cancellation between the gauge-dependent contributions to the running of the three-point vertex and the scalar anomalous dimension. Simultaneously, the gauge-independent contribution to those two expressions cancel as well, such that the beta function is given solely by the $\eta_A$ term
\bea
\beta_{\rho_3}\Big|_{\rm pert} &=& \rho_3 \frac{\eta_A}{2}= \frac{1}{48\pi^2} \rho_3^3.
\eea
As expected, the first equality is exactly what allows to read off the running of the coupling from the gauge field propagator.

To explicitly check that the FRG setup provides a unique 1-loop beta function for the gauge coupling irrespective of how it is read off from the right-hand-side of the Wetterich equation, we now calculate $\beta_{\rho_4}$.

\begin{figure}[!t]
\begin{center}
\includegraphics[width=0.22\linewidth,clip=true,trim=26cm 00cm 0cm 21cm]{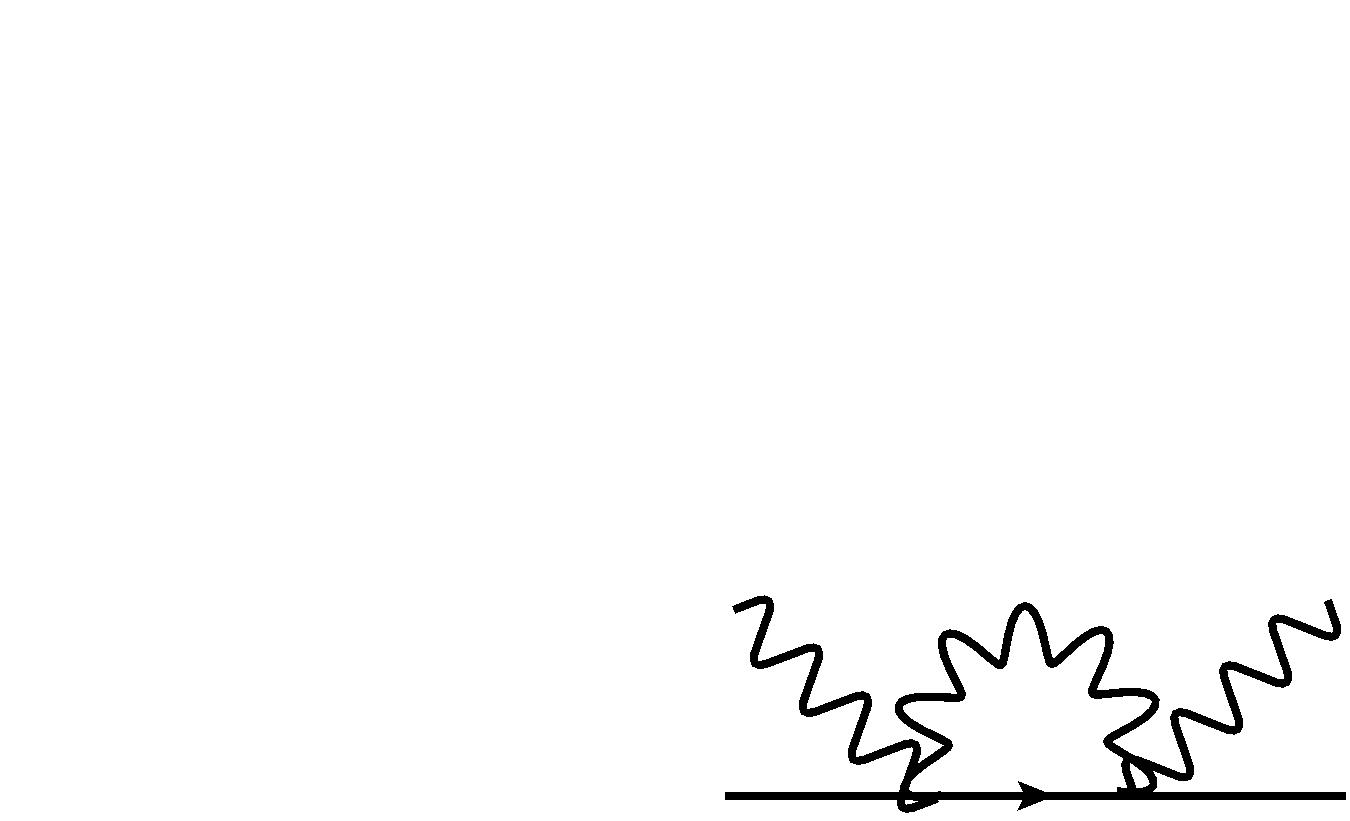}\quad\includegraphics[width=0.2\linewidth]{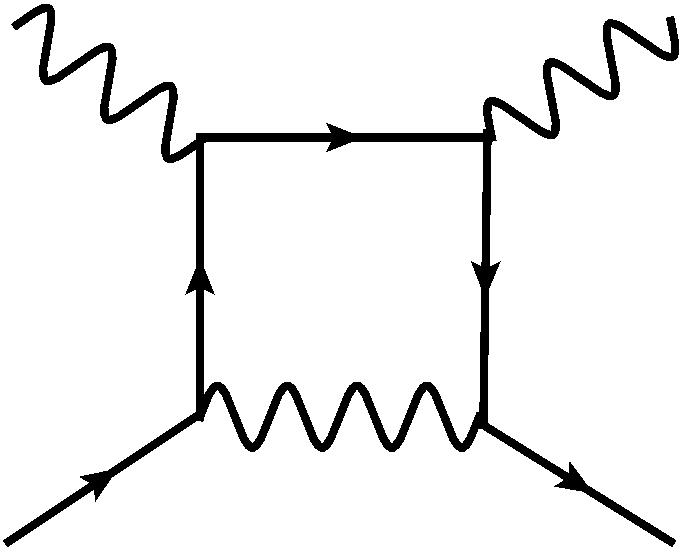}\quad\includegraphics[width=0.2\linewidth]{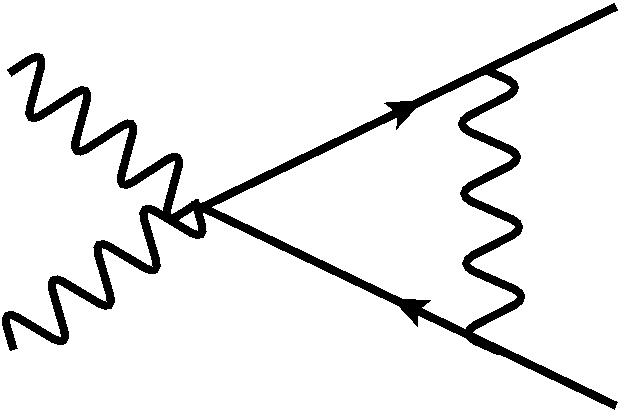}\quad
\includegraphics[width=0.2\linewidth]{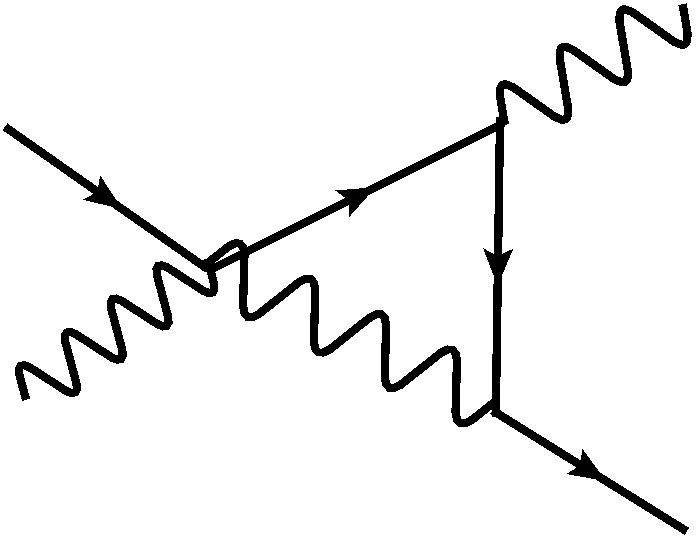}
\end{center}
\caption{\label{fig:betarho4mat} Four diagrams contribute to the running of the four-point vertex.}
\end{figure}

To project onto $\beta_{\rho_4}$, we use
\bea
\partial_t \rho_4 \nonumber&=&\Bigl(\frac{9}{76\rho_4}P_{\rho\mu}(q_4)P^{\mu}_{\sigma}(q_3)  
\frac{\delta^4 \partial_t \Gamma_k}{\delta\phi(q_2-q_3-q_4)\delta\phi^{\dagger}(q_2)\delta A_{\sigma}(q_3)\delta A_{\rho}(q_4)}
\Bigr)\Big|_{\phi^{\dagger}=\phi=A=0,\, q_i=0}.\label{eq:projrho4}
\eea
The separate contributions to the running of the four-point vertex, see Fig.~\ref{fig:betarho4mat} are gauge dependent,
\bea
\beta_{\rho_4}&=& \rho_4\left(\frac{\eta_A}{2}+ \frac{\eta_{\phi}}{2} \right) + \frac{(3+\xi)}{192\pi^2}\rho_4^3\left((6-\eta_A)+(6-\eta_{\phi}) \right)\\
&+&  \xi \Bigl(-\frac{8-\eta_A + 16 - 2\eta_{\phi}}{384\pi^2}\rho_3^2 \rho_4 - \frac{8 - \eta_A + 16 -2\eta_{\phi}}{192\pi^2} \rho_3^2 \rho_4+ \frac{10 - \eta_A + 30 - 3 \eta_{\phi}}{640\pi^2}\, \frac{\rho_3^4}{\rho_4} \Bigr).\nonumber
\eea

In the leading-order approximation the gauge-dependence cancels, leading to a gauge-independent perturbative result when we identify $\rho_3 =\rho_4=\rho$,
\be
\beta_{\rho_4} = \frac{\eta_A}{2}{\rho} = \frac{1}{48\pi^2}{\rho^3}.
\ee

Thus, the gauge-independence of the perturbative one-loop result is recovered from the functional Renormalization Group, despite the breaking of gauge invariance by the mass-like regulator. As expected, one-loop universality holds for dimensionless couplings also in this setting.

\section{Vertices}\label{App:vertices}
We use conventions in which $\phi(x) = \int d^4p\, \tilde{\phi}(p)e^{i\, p\cdot x}$, and drop the tilde on the Fourier transform throughout this work.
\be
\Gamma_k^{(2)}= \frac{\delta}{\delta \Phi(-p)}\frac{\delta}{\delta \Phi(q)}\Gamma_k,
\ee
where the field $\Phi$ summarizes all degrees of freedom and $\Phi(p) = (\phi(p), \phi^{\rm \dagger}(-p), A_{\mu}(p), h_{\mu\nu}(p))$. $\Gamma_k^{(2)}$ is a matrix in field space, the components of which we indicate by explicitly using the fields as indices, i.e., $\Gamma_{k\, \phi\, \phi^{\dagger}}(p,q) =  \frac{\delta}{\delta \phi(-p)}\frac{\delta}{\delta \phi^{\dagger}(-q)}\Gamma_k$.
In those conventions,

\bea
\Gamma_{k\, \phi\,  \phi^\dagger}(p,q)&=& \rho A_\mu(p-q)\, (q+p)^\mu+\rho^2\, \int_r\,  A_\mu(r)\, A^\mu(p-q-r),\\
\Gamma_{k\, \phi^\dagger\,  \phi}(p,q)&=&\Gamma_{k\, \phi\,  \phi^\dagger}(-q,-p),\\
\Gamma_{k\, A_\mu\,  \phi}(p,q)&=&\rho\,  (p-2q)^\mu\, \phi^\dagger(q-p) +2\rho^2\, \int_r\,  A^\mu(r+p-q)\, \phi^\dagger(r),\\
\Gamma_{k\, \phi\,  A_\mu }(p,q)&=&\Gamma_{k\, A_\mu\,  \phi}(-q,-p),\\
\Gamma_{k\, A_\mu\,  \phi^\dagger  }(p,q)&=&\rho\,  (2q-p)^\mu\, \phi(p-q)+2\rho^2\, \int_r\,  A^\mu(-r+p-q)\, \phi(r),\\
\Gamma_{k\, \phi^\dagger\,  A_\mu   }(p,q)&=&\Gamma_{k\, A_\mu\,  \phi^\dagger  }(-q,-p),\\
\Gamma_{k\,  A_\mu\,  A_\nu}(p,q)&=&2\rho^2\, \int_r\, \delta^{\mu\nu}\, \phi(r)\, \phi^\dagger(r+q-p),
\eea
 for the pure matter vertices. Applying a similar analysis to the vertices involving a graviton gives

\bea
&\Gamma&_{k\,  h_{\gamma\kappa}\, \phi}(p,q)\\
&=&\left[\frac{1}{2}\delta^{\gamma\kappa}\delta^{\mu\nu}-\delta^{\mu\left[\kappa\right.}\delta^{\left.\gamma\right]\nu} \right]\left( \phi^\dagger(q-p)\, q_\mu(q-p)_\nu\right. -\rho\, \int_r\, \phi^\dagger(r)\, (r+q)^\gamma\,  A^\kappa(r+p-q)\nonumber\\
&+&\left.\rho^2\, \int_{r\, m}\,  \phi^\dagger(r)\, A^\kappa(m)\, A^\gamma(r+p-m-q)  \right),\nonumber\\
&\Gamma&_{k\, \phi\,  h_{\gamma\kappa}}(p,q)= \Gamma_{k\,  h_{\gamma\kappa}\, \phi}(-q,-p),\\
&\Gamma&_{k\,  \phi^\dagger\,  h_{\gamma\kappa}}(p,q)\\
&=&\left[\frac{1}{2}\delta^{\gamma\kappa}\delta^{\mu\nu}-\delta^{\mu\left[\kappa\right.}\delta^{\left.\gamma\right]\nu} \right]\Bigl( \phi(p-q)\, p_\nu\, (p-q)_\mu  \Bigr.-\rho\, \int_r\, \phi(r)\, (r+p)^\gamma\,  A^\kappa(p-r-q)\,\nonumber \\
&+&\Bigl.\rho^2\, \int_{r\, m}\,  \phi^\dagger(r)\, A^\kappa(m)\, A^\gamma(p-m-r-q)\,   \Bigr),\nonumber\\
&\Gamma&_{k\,  h_{\gamma\kappa}\,  \phi^\dagger}(p,q)= \Gamma_{k\,  \phi^\dagger\,  h_{\gamma\kappa}}(-q,-p),\\
&\Gamma&_{k\, h_{\alpha\beta}\, h_{\gamma\kappa}}(p,q) = \frac{1}{2\alpha}\int_l\left( \delta^{\gamma(\alpha}\, \delta^{\beta)\kappa} - \frac{1}{2}\delta^{\alpha\beta}\, \delta^{\gamma\kappa}   \right)\, l^\mu\,  A_\mu(l)\, (p -q -l)^\mu\\
&+&  \frac{1}{2}\int_s\left[ \frac{1}{2}\delta^{\mu\nu}\, \delta^{\alpha\beta}\, \delta^{\gamma\kappa} - \delta^{\mu\nu}\, \delta^{\gamma(\alpha}\, \delta^{\beta)\kappa} + 4\delta^{(\mu(\alpha}\, \delta^{\beta)(\gamma}\, \delta^{\kappa)\nu)}\right.\left. - \delta^{\alpha\beta}\, \delta^{\mu(\gamma}\, \delta^{\kappa)\nu} - \delta^{\gamma\kappa}\, \delta^{\mu(\alpha}\, \delta^{\beta)\nu}  \right]\nonumber\\
&&\times \Bigl[ (s+p-q)_\mu\,  s_\nu\, \phi(s+p-q)\, \phi^\dagger(s)\Bigr.\Bigl. - \rho\int_r\,\Bigl( (r+s)_\mu\, \phi(r)\, \phi^\dagger(s)\, A_\nu(s+p-q-r) \nonumber\\
&&+ \rho\int_{l}\, A_\mu(l)\, A_\nu(s+p-q-r-l)\Bigr) \Bigr]\nonumber\\
&+&\frac{1}{2}\left[\delta^{\epsilon\lambda}\, \left( \delta^{\mu\nu}\left( \frac{1}{2}\delta^{\alpha\beta}\, \delta^{\gamma\kappa} - \delta^{\alpha(\gamma}\, \delta^{\kappa)\beta}   \right) + 4\delta^{(\mu(\alpha}\, \delta^{\beta)(\gamma}\, \delta^{\kappa)\nu)} \right.\right. \left.\left. - \delta^{\alpha\beta}\, \delta^{\mu(\gamma}\, \delta^{\kappa)\nu)} - \delta^{\gamma\kappa}\, \delta^{\mu(\alpha}\, \delta^{\beta)\nu)} \right) \right.\nonumber\\
&+& \left. \delta^{\mu\nu}\left(  4\delta^{(\epsilon(\alpha}\, \delta^{\beta)(\gamma}\, \delta^{\kappa)\lambda)} - \delta^{\alpha\beta}\, \delta^{\epsilon(\gamma}\, \delta^{\kappa)\lambda)} - \delta^{\gamma\kappa}\, \delta^{\epsilon(\alpha}\, \delta^{\beta)\lambda)} \right)\right.\nonumber\\
&+&\left.  2\delta^{\mu(\alpha}\, \delta^{\beta)\nu}\, \delta^{\epsilon(\gamma}\, \delta^{\kappa)\lambda} 2\delta^{\mu(\gamma}\, \delta^{\kappa)\nu}\, \delta^{\epsilon(\alpha}\, \delta^{\beta)\lambda} \right]\nonumber\\
&& \times \int_l \left( 2l_{(\mu}\, A_{\epsilon)}(l)\, (p-l-q)_\lambda\, A_\nu(p-l-q) - 2l_{(\mu}\, A_{\epsilon)}(l)\, (p-l-q)_\nu\, A_\lambda(p-l-q)  \right).\nonumber
\eea

For the projections onto the couplings, we work with particular choices for the external momenta, choosing them totally symmetric for the projection onto $\rho_3$, such that
\bea
   &{}& q_1=\left| q \right|\, \begin{pmatrix}
         \frac{1}{2}\\
         \frac{\sqrt{3}}{2}\\
         0\\
         0\\
        \end{pmatrix}\quad,\quad
    q_2=\left| q \right|\, \begin{pmatrix}
         1\\
         0\\
         0\\
         0\\
        \end{pmatrix}\quad,\quad
   q_3=\left| q \right|\, \begin{pmatrix}
         \frac{1}{2}\\
         -\frac{\sqrt{3}}{2}\\
         0\\
         0\\
        \end{pmatrix}.\nonumber
\eea
Similarly, the symmetric projection onto $\rho_4$ can be derived
\bea
   &{}& q_1=\left| q \right|\, \begin{pmatrix}
         1\\
         0\\
         0\\
         0\\
        \end{pmatrix}\quad,\quad
    q_2=\left| q \right|\, \begin{pmatrix}
         \frac{1}{3}\\
         -2\frac{\sqrt{2}}{3}\\
         0\\
         0\\
        \end{pmatrix}\quad,\quad
  q_3=\left| q \right|\, \begin{pmatrix}
         -\frac{1}{3}\\
         -\frac{\sqrt{2}}{3}\\
         -\sqrt{\frac{2}{3}}\\
         0\\
        \end{pmatrix}\quad, \quad\nonumber
    q_4=\left| q \right|\, \begin{pmatrix}
         -\frac{1}{3}\\
         -\frac{\sqrt{2}}{3}\\
         \sqrt{\frac{2}{3}}\\
         0\\
        \end{pmatrix}.\nonumber
\eea
For the TT approximation, we project out all the modes of the graviton propagator with exception of the transverse, traceless mode. This can be accomplished by using the following projector
\bea
\label{ttaprox}
P^{TT}_{\alpha\beta\mu\nu}(p)=\frac{1}{2} P_{\alpha\nu}\,  P_{\beta\mu}+\frac{1}{2} P_{\alpha\mu}\,  P_{\beta\nu}-\frac{1}{3} P_{\alpha\beta}\, P_{\mu\nu},
\eea
where $p$ is the internal loop momentum running through the graviton propagator and we use the transverse projector
\be
P_{\alpha\beta}(p) =\left(\delta_{\alpha\beta} - \frac{p_\alpha\,  p_\beta}{p^2} \right).
\ee

\end{appendix}

\end{document}